\documentclass[aps,prb,twocolumn,superscriptaddress,floatfix,showpacs,10pt,letterpaper]{revtex4}
\usepackage{graphicx}
\usepackage{dcolumn}
\usepackage{bm}
\usepackage{subfigure}
\usepackage[USenglish]{babel}
\usepackage{amssymb,amsmath}
\usepackage{relsize}
\usepackage{lmodern}
\usepackage{slantsc}
\usepackage{scalefnt}
\usepackage{tikz}
\usetikzlibrary{arrows,calc,scopes}
\usetikzlibrary{shapes.geometric}
\allowdisplaybreaks[1]

\usepackage[colorlinks,citecolor=blue]{hyperref}

\newcommand{\be}{\begin{equation}} \newcommand{\ee}{\end{equation}}
\newcommand{\bse}{\begin{subequations}}\newcommand{\ese}{\end{subequations}}
\newcommand{\bpm}{\begin{pmatrix}} \newcommand{\epm}{\end{pmatrix}}
\newcommand{\bmm}{\begin{matrix}} \newcommand{\emm}{\end{matrix}}

\newcommand{\Z}{\mathbb{Z}} 
\newcommand{\R}{\mathbb{R}}

 \renewcommand{\t}[1]{\tilde{#1}}

\newcommand{\e}{\hspace{1pt}\mathrm{e}}
\newcommand{\dd}{\hspace{1pt}\mathrm{d}}
\newcommand{\imth}{\hspace{1pt}\mathrm{i}\hspace{1pt}}

\newcommand{\Rf}[1]{Ref.~\onlinecite{#1}}

\newcommand{\eq}[1]{(\ref{#1})} \newcommand{\eqn}[1]{eqn.~(\ref{#1})}

\newcommand{\<}{\langle} \renewcommand{\>}{\rangle}
 
\newcommand{\Tr}{{\rm Tr}}    
 \newcommand{\prt}{\partial}

\newcommand{\ie}{{\it ie~}}  \newcommand{\etc}{{\it
etc~}}

\newcommand{\al}{\alpha} \newcommand{\bt}{\beta} 
 \newcommand{\eps}{\epsilon}
 \newcommand{\ga}{\gamma}
\newcommand{\Ga}{\Gamma}  \newcommand{\la}{\lambda}
 \newcommand{\om}{\omega} 
\renewcommand{\th}{\theta}  \newcommand{\si}{\sigma}

 \newcommand{\cH}{ {\cal H} }  \newcommand{\cL}{ {\cal L} }

\newcommand{\Blangle}{\Biggl\langle\bmm\vspace{-3pt}
  \scalefont{0.6}} 
\newcommand{\BRvert}{\emm\Biggr\vert} 
\newcommand{\Bvert}{\emm\Biggr\vert\bmm\vspace{-3pt}\scalefont{0.6}} 
\newcommand{\Brangle}{\emm\Biggr\rangle}

\newcommand{\sline}{\begin{tikzpicture}[scale=0.6,baseline]
\draw (0,0) -- (0.5,0) node[above,inner sep=0, outer sep=1] {\scalebox{0.8}{$i$}} -- (1,0);
\draw[->,>=stealth',line width=0.01pt] (0.4,0) -- (0.5,0);
\end{tikzpicture}}

\newcommand{\curveline}{\begin{tikzpicture}[scale=0.8,baseline]
\draw (0,0) -- ++(0.25,0) -- ++(300:0.25) -- ++(0.25,0) -- ++(60:0.25) -- +(0.25,0);
\node[inner sep=0,outer sep=1] at(0.5,0) {\scalebox{0.8}{$i$}};
\draw[->,>=stealth',line width=0.01pt] (0,0) ++(0.25,0) ++(300:0.25) -- ++(0.15,0);
\end{tikzpicture}}

\newcommand{\Ygraph}[4][1]{
\begin{tikzpicture}[scale=0.6,baseline]
  \draw [->,>=stealth',line width=0.01pt] (30:0.1) -- (0,0) ; 
    \draw (30:1) -- (0,0) ; 
  \draw [->,>=stealth',line width=0.01pt] (150:0.1) -- (0,0); 
    \draw (150:1) -- (0,0); 
  \node at(-0.5,0.01) {$#2$};
  \node at(0.5,0.01) {$#4$};
  \ifnum #1=1 {
   \draw [->,>=stealth',line width=0.01pt] (0,-1/20) -- (0,0); 
   \node at(-0.2,-0.5) {$#3$};
  }
  \else{
    \draw [<-,>=stealth',line width=0.01pt] (0,-1/2) -- (0,0); 
    \node at(-0.4,-0.5) {$-#3$};
    }
  \fi
    \draw (0,-1) -- (0,0); 
  \end{tikzpicture}
  }

\newcommand{\Hgraph}[3][1]{
  \begin{tikzpicture}[scale=0.6]
    \coordinate (c) at (0,0);
    \coordinate (l) at (-0.7, 0);
    \coordinate (r) at ($ (c) ! -1 ! (l) $);
    \coordinate (ul) at (-0.95,0.75);
    \coordinate (lr) at ($ (c) ! -1 ! (ul) $);
    \coordinate (ll) at (-0.95,-0.75);
    \coordinate (ur) at ($ (c) ! -1 ! (ll) $);
    \draw (l) -- (r) ; 
    \draw (ul) -- (l); 
    \draw (ll) -- (l); 
    \draw (lr) -- (r); 
    \draw (ur) -- (r); 
    \ifnum #1=1 {
    \node[right] at($ (ul) ! .3 ! (l) $) {$#2_{\text{\scalebox{0.7}{$1$}}} $};
    \node[right] at($ (ll) ! .2 ! (l) $) {$#2_{\text{\scalebox{0.7}{$2$}}} $};
    \node[left] at($ (lr) ! .2 ! (r) $) {$#2_{\text{\scalebox{0.7}{$3$}}} $};
    \node[left] at($ (ur) ! .3 ! (r) $) {$#2_{\text{\scalebox{0.7}{$4$}}} $};
    \node[below] at($ (c) ! .15 ! (0,0.5) $) {$#3 $};}
    \fi
    \draw[<-,>=stealth', line width=0.01pt] (l) -- (r) ;
    \draw[->,>=stealth', line width=0.01pt] (ul) -- (l);
    \draw[->,>=stealth', line width=0.01pt] (ll) -- (l);
    \draw[->,>=stealth', line width=0.01pt] (lr) -- (r);
    \draw[->,>=stealth', line width=0.01pt] (ur) -- (r);
   \end{tikzpicture}
  }

 \newcommand{\Xgraph}[3][1]{
  \begin{tikzpicture}[scale=0.6]
    \coordinate (c) at (0,0);
    \coordinate (u) at (0, 0.6);
    \coordinate (d) at ($ (c) ! -1 ! (u) $);
    \coordinate (ul) at (-0.85,0.85);
    \coordinate (lr) at ($ (c) ! -1 ! (ul) $);
    \coordinate (ll) at (-0.85,-0.85);
    \coordinate (ur) at ($ (c) ! -1 ! (ll) $);
    \draw (u) -- (d) ; 
    \draw (ul) -- (u); 
    \draw (ll) -- (d); 
    \draw (lr) -- (d); 
    \draw (ur) -- (u); 
    \ifnum #1=1 {
    \node[below] at($ (ul) ! .3 ! (u) $) {$#2_{\text{\scalebox{0.7}{$1$}}} $};
    \node[above] at($ (ll) ! .2 ! (d) $) {$#2_{\text{\scalebox{0.7}{$2$}}} $};
    \node[above] at($ (lr) ! .2 ! (d) $) {$#2_{\text{\scalebox{0.7}{$3$}}} $};
    \node[below] at($ (ur) ! .2 ! (u) $) {$#2_{\text{\scalebox{0.7}{$4$}}} $};
    \node[right] at($ (-0.5,0) ! 0.8 ! (c) $) {$#3 $}; }
    \fi
    \draw[<-,>=stealth', line width=0.01pt] (u) -- (d) ;
    \draw[->,>=stealth', line width=0.01pt] (ul) -- (u);
    \draw[->,>=stealth', line width=0.01pt] (ll) -- (d);
    \draw[->,>=stealth', line width=0.01pt] (lr) -- (d);
    \draw[->,>=stealth', line width=0.01pt] (ur) -- (u);
   \end{tikzpicture}
  }

\newcommand{\Psix}[3][1]{
\begin{tikzpicture}[scale=0.8]
\node[name=s, regular polygon, regular polygon sides=6, minimum size=1cm, outer sep=0pt ,draw] at (0,0) {}; 
\foreach \anchor/\x/\y /\xx/\yy /\b in
{corner 1/0.17/0.17*1.732/-0.11/0.18/1, corner 2/-0.17/0.17*1.732/0.07/0.18/2, corner 3/-0.34/0/-0.15/-0.18/3, corner 4/-0.17/-0.17*1.732/-0.22/-0.05/4, corner 5/0.17/-0.17*1.732/0.2/-0.05/5, corner 6/0.34/0/0.15/-0.18/6}
{
 \draw[shift=(s.\anchor)] (0,0) -- (\x,\y) node at(\xx,\yy) {$#2_{\text{\scalebox{0.7}{$\b$}}}$};
 \ifnum #1=1
 \draw[shift=(s.\anchor),<-,>=stealth', line width=0.01pt] (s.\anchor) -- (\x,\y);
 \fi
 }
\foreach \anchor/\xx/\yy /\a in
{side 1/0/-0.18/1, side 2/-0.18/0.05/2, side 3/0.15/0.05/3, side 4/0/-0.18/4, side 5/-0.18/0.05/5, side 6/0.15/0.05/6}
 \draw[shift=(s.\anchor)]  node at(\xx,\yy) {$#3_{\text{\scalebox{0.7}{$\a$}}}$};
\ifnum #1=1{
  \foreach \anchorr/\anchorf in
   {corner 1/corner 2, corner 2/corner 3, corner 3/corner 4, corner 4/corner 5, corner 5/corner 6, corner 6/corner 1}
   \draw[shift=(s.\anchorr), ->, >=stealth', line width=0.01pt]  (s.\anchorr) -- (s.\anchorf);}
 \else {
  \foreach \anchorb/\anchorw in
   {corner 1/corner 2, corner 3/corner 4, corner 5/corner 6} {
   \node[fill=black, circle, minimum size=2.5, inner sep=0, outer sep=0, draw] at(s.\anchorb) {};
   \node[fill=white, circle, minimum size=2.5, inner sep=0, outer sep=0, draw] at(s.\anchorw) {};}
}
\fi
\end{tikzpicture}
}

\newcommand{\bubbleGraph}[4]{\begin{tikzpicture}[scale=0.6]
  \draw (1,0) arc (0:180:0.5 and 0.4) ;
  \draw [->,>=stealth', line width=0.01pt] (1,0) arc (0:90:0.5 and 0.4) node[below left] {\scalebox{0.5}{$#3$}};
  \draw (0,0) arc (180:360:0.5 and 0.4);
  \draw [->,>=stealth', line width=0.01pt] (0,0) arc (180:270:0.5 and 0.4) node[above right] {\scalebox{0.5}{$#4$}};
  \draw (-0.5,0) -- (0,0);
  \draw [->,>=stealth', line width=0.01pt] (-0.5,0) -- (-0.2,0) node[above, inner sep=0,outer sep=1] {\scalebox{0.5}{$#1$}};
  \draw (1,0) -- (1.5,0);
  \draw [->,>=stealth', line width=0.01pt] (1,0) -- (1.35,0) node[above, inner sep=0,outer sep=1] {\scalebox{0.5}{$#2$}};
  \end{tikzpicture}}

\newcommand{\loopGraph}[1]{\begin{tikzpicture}[scale=0.4,baseline]
 \draw (1,0) arc (0:360:0.5) node[left] {\scalebox{0.8}{$#1$}};
 \draw[->, >=stealth',line width=0.01pt] (1,0) arc (0:10:0.5); \end{tikzpicture}}

\newcommand{\shadowGraph}{\begin{tikzpicture}[scale=0.5,baseline]
 \draw[fill=gray,rounded corners,ultra thick, gray] (-0.4,0.5) rectangle (0.4,-0.5);
 \end{tikzpicture}}

\newcommand{\honeycomb}[2]{\begin{tikzpicture}[scale=0.5]
   \clip[draw] (0,0) rectangle (2*1.732,-3);
\def\hexagonpath{ +(30:1) -- +(90:1)  -- +(150:1) -- +(210:1) -- +(270:1) -- +(330:1)  -- cycle }
\foreach \x in {0,...,#1}
  \foreach \y in {0,...,#2} {
    \ifodd\x
     \draw (0,0) ++(0,-1/2-3*\x/2) ++(30:1) ++(30:\y * 2) ++(0,-\y) \hexagonpath;
    \else\draw (0,0) ++(0,-3*\x/2) ++(30:\y * 2) ++(0,-\y) \hexagonpath;
    \fi }
\end{tikzpicture}}

\usepackage{varioref}

\begin{document}


\begin{titlepage}

\title{Quantized topological terms in weakly coupled gauge theories\\
and their connection to symmetry protected topological phases}

\author{Ling-Yan Hung}\email{lhung@physics.harvard.edu}
\affiliation{Perimeter Institute for Theoretical Physics, 31 Caroline St N,
Waterloo, ON N2L 2Y5, Canada}
\affiliation{Department of Physics, Harvard University, Cambridge MA 02138}
\author{Xiao-Gang Wen}
\affiliation{Perimeter Institute for Theoretical Physics, 31 Caroline St N,
Waterloo, ON N2L 2Y5, Canada}
\affiliation{Department of Physics, Massachusetts Institute of Technology,
Cambridge, Massachusetts 02139, USA}
\affiliation{Institute for Advanced Study, Tsinghua University, Beijing,
100084, P. R. China}

\date{\today}

\begin{abstract}

We consider a weakly coupled gauge theory where charged particles all have large gaps
(\ie no  Higgs condensation to break the gauge ``symmetry'') and the field
strength fluctuates only weakly.  We ask, what kind of topological terms (the terms
that do not depend on space-time metrics) can be added to the Lagrangian of
such a weakly coupled gauge theory.  For example, for weakly coupled
$U(1)$ gauge theory in $d=3$
space-time dimensions, a Chern-Simons topological term $\frac{k}{4\pi}
a_\mu\prt_\nu a_\la \eps^{\mu\nu\la},\ k\in \Z$ can be added.

In this paper, we systematically construct quantized topological terms which
are generalization of the Chern-Simons terms and $F\wedge F$ terms, in any
space-time dimensions and for any gauge groups (continuous or discrete).  We
can use each element of the topological cohomology classes $H^{d+1}(BG,\Z)$ on the
classifying space $BG$ of the gauge group $G$ to construct a quantized
topological term in $d$ space-time dimensions.

In 3$d$ or for finite gauge groups above 3$d$, the weakly coupled gauge theories are
gapped.  So our results on topological terms can be viewed as a systematic
construction of gapped topologically ordered phases of weakly coupled gauge theories.  In
other cases, the weakly coupled gauge theories are gapless.  So our results can be viewed
as an attempt to systematically construct different gapless phases of weakly coupled
gauge theories.

Amazingly, the bosonic symmetry protected topological (SPT) phases with a
finite on-site symmetry group $G$ are
also classified by the same $H^{d+1}(BG,\Z)$.
(SPT phases are gapped quantum phases with a symmetry and \emph{trivial}
topological order.) In this paper, we show an explicit duality relation between
topological gauge theories with the quantized topological terms and the bosonic
SPT phases, for any finite group $G$ and in any dimensions; a
result first obtained by Levin and Gu.  We also study the relation between
topological lattice gauge theory and the string-net states with non-trivial
topological order and no symmetry.

\end{abstract}

\maketitle

\end{titlepage}

{\small \setcounter{tocdepth}{1} \tableofcontents }

\section{Introduction}

Quantum many-body systems can be described by Lagrangians in space-time.  As we
change the coupling constants in a Lagrangian a little, the new Lagrangian will
describe a new quantum many-body system which is usually in the same phase as
the original system.  However, the  Lagrangians may also contain the so called
topological terms -- the terms that do not depend on space-time metrics.  Some
of those topological terms are quantized. Adding quantized topological terms to
a Lagrangian will generate new Lagrangians that usually describe different
phases of quantized many-body systems.

So studying and classifying quantized topological terms is one way to
understand and classify different possible phases of quantum many-body systems.
For example, a non-linear $\si$-model
\begin{align}
 \cL= \frac{1}{\la} [\prt g(x^\mu)]^2, \ \ g\in G
\end{align}
with symmetry group $G$ can be in a disordered phase that does not break the
symmetry $G$ when $\la$ is large.  By adding different $2\pi$ quantized
topological $\th$-terms to the Lagrangian $\cL$, we can get different
Lagrangians that describe different disordered phases that does not break the
symmetry $G$.\cite{CGL1172}  Those disordered phases are the symmetry protected
topological (SPT) phases.\cite{GW0931,PBT0959} We find that the different
$2\pi$ quantized topological $\th$-terms in $d$-dimensional space-time can be
classified by the Borel group cohomology classes $\cH^d_B(G,\R/\Z)$, which leads
us to believe that the SPT phases in $d$-dimensional space-time can be
classified by $\cH^d_B(G,\R/\Z)$.\cite{CLW1141,CGL1172} So the possible
topological terms in a non-linear $\si$-model help us to understand the
possible phases of the non-linear $\si$-model.

This motivated us to study possible quantized topological terms in gauge
theory, hoping to understand different phases of gauge theory in a more
systematic way.  But a gauge theory is a very complicated system whose low
energy phases can be any thing. To make our problem better defined, we need to
restrict ourselves to a special class of gauge theories that we will call
``weakly coupled gauge theories'' or ``weakly coupled lattice gauge theories''.
``Weakly coupled (lattice)
gauge theories'', by definition, are lattice gauge theories in the weak coupling
limit (where gauge flux through each plaquette is small) and all the gauge
charged excitations have a large energy gap (\ie no Higgs condensations).

We find that a quantized topological term in the weakly coupled gauge theories can be
constructed from each element in topological cohomology class
$H^{d+1}(BG,\Z)$ for the classifying space $BG$ of the gauge group $G$
in $d$ space-time dimensions.

The result is obtained by applying the three approaches introduced by Baker and
by Dijkgraaf and Witten\cite{B7741,DW9093} to lattice gauge theories.

In the first approach (see section II),
we follow Dijkgraaf and Witten\cite{DW9093} to use
cocycles in group cohomology class $\cH^d_B(G,\R/\Z)$ to construct topological
terms in weakly coupled gauge theories in $d$-dimensional space-time with a finite gauge
group $G$.  Such weakly coupled gauge theories are gapped and describe topological
phases.  So we can use $\cH^d_B(G,\R/\Z)$ to describe different topological
phases in  weakly coupled gauge theories with finite gauge group.

In the second approach (see section III), we view a lattice gauge theory as a
lattice non-linear $\si$-model whose target space is given by the classifying
space $BG$ of the gauge group $G$.  So the quantized topological terms in a
weakly coupled lattice gauge theory can be described by the quantized topological terms
in the non-linear $\si$-model of the classifying space $BG$.  Such a quantized
topological term is similar to the one used in the classification of SPT
phases,\cite{CGL1172,GW1248} where the target space is simply the symmetry
group $G$.  Using this approach, we find that some quantized topological terms
in weakly coupled gauge theories with gauge group $G$ can be constructed from the torsion
elements in topological cohomology class $\text{Tor}[H^{d+1}(BG,\Z)]$ of the
classifying space $BG$. Since  $\cH_B^{d}(G, \R/\Z) =
H^{d+1}(BG,\Z)=\text{Tor}[H^{d+1}(BG,\Z)]$ for finite groups,\cite{CGL1172} the
second approach contains the result of the first approach.

In the third approach (see section IV), we express $d$-dimensional Chern-Simons
terms (assuming $d$ = odd) in terms of the differential characters of Chern,
Simons and Cheeger.\cite{B7741,DW9093}
The different  differential characters are classified by $H^{d+1}(BG,
\Z)$,\cite{B7741} thus in turn giving a classification of Chern-Simons terms
with gauge group $G$.  We see that the quantized topological terms obtained
from the third approach (classified by $H^{d+1}(BG,\Z)$) contain those obtained
from the second approach (classified by $\text{Tor}[H^{d+1}(BG,\Z)]$), in $d$ =
odd space-time dimensions.

Also, in $d=$ even space-time dimensions,
$H^{d+1}(BG,\Z)=\text{Tor}[H^{d+1}(BG,\Z)]$ (see \eqn{HTorH}).  So  the
quantized topological terms are classified by $H^{d+1}(BG,\Z)$ in any
dimensions.  Since
\begin{align}
\cH_B^{d}(G, \R/\Z)\simeq H^{d+1}(BG, \Z)
\end{align}
[see eqn. (J32) in \Rf{CGL1172}], this suggests that the quantized topological
terms in weakly coupled gauge theories are described by $\cH_B^{d}(G, \R/\Z)$. Here $G$
can be a continuous group.  Thus the quantized topological terms is the dual of
the SPT phases classified by the same $\cH^d_B(G,\R/\Z)$.

We know that, for continuous gauge group $G$, a kind of quantized topological
terms -- Chern-Simons terms --  can be defined in any $d$ = odd dimensions:
\begin{align}
S_{3d}& =\int_M \frac{2\pi K}{2! (2\pi)^2} \Tr(AF-\frac{A^3}{3})
+\frac{F_{\mu\nu}F^{\mu\nu}}{\la},
\nonumber\\
S_{5d}& =\int_M \frac{2\pi K}{3! (2\pi)^3} \Tr(AF^2-\frac{A^3F}{2}
-\frac{A^5}{10}) +\frac{F_{\mu\nu}F^{\mu\nu}}{\la},
\nonumber\\
... ...
\end{align}
where $K$ = integer, $A$ is the gauge potential one form and $F$ the gauge
field strength two form.
Also, for example, $AF^2$ means the wedge product $A\wedge F\wedge F$.
We have also included the usual kinetic term
in the above with coefficient $1/\lambda$.
In $d=3$ space-time dimensions, a  Chern-Simons gauge
theory is gapped and describes a topologically ordered phase.  So the
topological phases of a $d=3$ weakly coupled gauge theory are described by
$\cH_B^3(G,\R/\Z)$.  Beyond $d=3$, the above  Chern-Simons gauge theory is
gapless for small $\la$, and the Chern-Simons term is irrelevant at low
energies.  However, the Chern-Simons term is not renormalized, and we believe
that different Chern-Simons terms  will describe different gapless phases of
the weakly coupled gauge theory.  Consider $d=5$ $U(1)$ Chern-Simons theory, if the
space-time has a topology of $M^3\times S^2$ and the $U(1)$ gauge field has a
non-zero total flux through $S^2$, the  $d=5$ $U(1)$ Chern-Simons theory on
$M^3\times S^2$ will becomes a non-trivial  $d=3$ $U(1)$ Chern-Simons theory on
$M^3$ which describes a non-trivial topological phase.

The above discussion is for $d$ = odd dimensions. Similarly, in $d$ = even
space-time dimensions, we can have the following weakly coupled gauge theories
with topological term
\begin{align}
S_{4d}& =\int_M \frac{\th}{2! (2\pi)^2} \Tr(F^2)
+\frac{F_{\mu\nu}F^{\mu\nu}}{\la},
\nonumber\\
S_{6d}& =\int_M \frac{\th}{3! (2\pi)^3} \Tr(F^3) +\frac{F_{\mu\nu}F^{\mu\nu}}{\la},
\nonumber\\
... ...
\end{align}
where $\th \in [0,2\pi)$ and $\la$ is small.

The above topological terms are  not quantized and do not give rise to new
phases.  In this paper, we show  that quantized  topological terms do exist for
weakly coupled gauge theories in $d=$ even space-time dimensions.  Such quantized
topological terms are described by $H^{d+1}(BG,\Z)=$ Tor$[H^{d+1}(BG,\Z)]$.
Again, the quantized topological terms do not open a mass gap in $d\geq 6$
dimensional space-time (for small $\la$).  Here we like to propose that the
different quantized topological terms give rise to different gapless phases of
the gauge theories.  The situation is even more interesting (and unclear) in
$d=4$ dimensional space-time, where some gauge theories are confined even in
small $\la$ limit.  It is not clear if  the different quantized topological
terms give rise to different confined phases.

Since the introduction of topological order in 1989,\cite{Wtop,Wrig} we have
introduced many ways of constructing topologically ordered states (\ie long
range entangled states) on lattice: resonating-valence-bond
state,\cite{RK8876,MS0312} projective
construction,\cite{BZA8773,AZH8845,DFM8826,WWZcsp,RS9173,Wsrvb,W9927}
string-net condensation,\cite{FNS0428,LW0510,CGW1038,GWW1017} weakly coupled gauge
theory of finite gauge group, \etc.

The usual construction of weakly coupled lattice gauge theories for a given finite gauge
group only gives rise to one type of topological order.  In this paper, we
manage to ``twist'' the weakly coupled gauge theory with a given finite gauge group by
adding topological terms to obtain more general topological orders.  The added
topological terms are constructed from
$\cH_B^{d}(G,\R/\Z)=H^{d+1}(BG,\Z)$ (see table \ref{tb}).

\begin{table}[tb]
 \centering
 \begin{tabular}{ |c||c|c|c|c| }
 \hline
 group $G$ & $d=1$ & $d=2$ & $d=3$ & $d=4$  \\
\hline
\hline
$Z_n$ & $\Z_n$  & $0$ & $\Z_n$ & $0$    \\
$Z_m \times Z_n$ & $\Z_m\oplus \Z_n$ & $\Z_{(m,n)}$ & {\scriptsize $\Z_m\oplus \Z_n\oplus \Z_{(m,n)}$} & $ \Z^2_{(m,n)}$   \\
$Z_n \rtimes Z_2$ & $\Z_2\oplus \Z_{(2,n)}$ & $\Z_{(2,n)}$ & {\scriptsize $\Z_n\oplus \Z_2\oplus \Z_{(2,n)}$} & $\Z^2_{(2,n)}$   \\
\hline
{$U(1)$} & {$\Z$}  & {$0$} & {$\Z$} & {$0$}    \\
$U(1) \rtimes Z_2$ & $\Z_2$ & $\Z_2$ & $\Z\oplus \Z_2$ & $\Z_2$   \\
$U(1) \times Z_2$ & $\Z\oplus \Z_2$ & $0$ & $\Z\oplus \Z^2_2$ & $0$   \\
{\scriptsize $U(1)\times U(1) \rtimes Z_2$} & $\Z_2\oplus \Z$ & $\Z_2$ & $\Z^2\oplus \Z_2^2$ & $\Z_2^2$   \\
\hline
$SU(2)$ & $0$  & $0$ & $\Z$ & $0$    \\
$SO(3)$ & $0$  & $\Z_2$ & $\Z$ & $0$    \\
\hline
 \end{tabular}
 \caption{
The above table lists $\cH_B^{d}[G,\R/\Z]=H^{d+1}[G,\Z]$ whose elements label
the quantized topological terms of weakly coupled lattice gauge theories with gauge
group $G$ in $d$ space-time dimensions.  Here $0$ means that our construction
only gives rise to a trivial topological term which is zero.  $\Z_n$ means that
the constructed non-trivial topological terms plus the trivial one are labeled
by the elements in $\Z_n$.  $\Z_n^2$ means $\Z_n\oplus \Z_n$.  $U(1)$
represents $U(1)$ symmetry, $Z_n$ represents cyclic symmetry, \etc.  Also
$(m,n)$ is the greatest common divisor of $m$ and $n$.
}
 \label{tb}
\end{table}

We also manage to ``twist'' the weakly coupled gauge theory of continuous gauge group  by
adding topological terms.  The added  topological terms in this case are
constructed from $H^{d+1}(BG,\Z)=\cH_B^{d}(G,\R/\Z)$.  This leads to more
general topological orders in (2+1)D than those described by the standard
Chern-Simons terms.  But in higher dimensions, this may lead to  more general
gapless phases of weakly coupled gauge theories (see table \ref{tb}).

From \Rf{R1182}, we find that $H^7(U(1)\rtimes Z_2, \Z)=\Z_2\oplus \Z_2$.  So
there are three non-trivial quantized topological terms that we can add to the
$U(1)\rtimes Z_2$ weakly coupled gauge theory in $6$-dimensional space-time.  We think
that one of the quantized topological term is given by
\begin{align}
S_{U(1)\rtimes Z_2}& =\int_M \frac{\th}{3! (2\pi)^3} F\wedge F \wedge F
+\frac{(F_{\mu\nu})^2}{\la}, \ \
\th=\pi
\end{align}
where $F$ is the field strength of the $U(1)$ gauge field and the $Z_2$ gauge
transformation acts as $F\to -F$.\cite{BLP9203}  Such a $U(1)\rtimes Z_2$ gauge
theory can be viewed as the  $U(1)$ gauge theory with the $Z_2$ gauged charge
conjugation symmetry.  Note that the theory has a charge conjugation symmetry
only when $\th=0,\pi$.

The gapless phases (in small $\la$ limit) for $\th=0$ and $\th=\pi$ should be
different.

Similarly, we can have a quantized topological term in $U(1)\times U(1)\rtimes
Z_2$ weakly coupled gauge theory in $4$-dimensional space-time:
\begin{align}
S_{U(1)\times U(1)\rtimes Z_2}&
=\int_M \frac{\th}{(2\pi)^2} \t F \wedge F
+\frac{(F_{\mu\nu})^2}{\la}
+\frac{(\t F_{\mu\nu})^2}{\la},
\end{align}
where $\th=\pi$, $F,\t F$ are the field strengthes of the two $U(1)$ gauge
fields and the $Z_2$ gauge transformation acts as $F\to -F$.  In such a
$U(1)\times U(1)\rtimes Z_2$ gauge theory, the $U(1)\times U(1)$ magnetic
monopole with monopole charge $(m,\t m)$ will carry electric charges $(q,\t
q)=( \frac{\t m\th}{2\pi}+\text{int.}, \frac{
m\th}{2\pi}+\text{int.})$.\cite{W7983,RF1005} We see that when $\th=\pi$, dyons
with quantum numbers $(q,m;\t q, \t m)=( \pm \frac12,\pm 1; \pm \frac12,\pm 1)$
all appear, which is consistent with the $Z_2$ gauge symmetry.

When $\th=0$, a dyon with quantum number $(q,m;\t q, \t m)$ has a statistics
$(-)^{qm+\t q \t m}$ (where $+$ means Bose statistics and $-$ means Fermi
statistics and note that $q$, $\t q$, $m$ and $\t m$ are all
integers).\cite{T3141,JR7616,W8246,G8205,LM0012} It is shown in \Rf{GMW8921}
that the statistics of a  dyon in the presence of the $F\wedge \t F$ term does
not depend on $\th$. So for a non-zero $\th$, the satistics is given by
\begin{align}
(-)^{(q-\frac{\t m\th}{2\pi}) m+(\t q-\frac{m \th}{2\pi}) \t m}
=(-)^{q m+\t q \t m -\frac{\th}{\pi} m \t m}
\end{align}
where the electric charges $q,\ \t q$ are no longer integers: $q=\frac{\t
m\th}{2\pi}$ + integer and $\t q=\frac{m\th}{2\pi}$ + integer.  We see that
when $\th=\pi$, the statistics is invariant under the $Z_2$ gauge
transformation $(q,m;\t q, \t m) \to (-q,-m;\t q, \t m)$.  A $(q,m;\t q, \t
m)=(\frac12,1;\frac12,1)$ dyon is a boson while a $(q,m;\t q, \t
m)=(\frac12,1;-\frac12,1)$ dyon is a fermion.  Clearly, the $\th=0$ and
$\th=\pi$ correspond to two different gapless phases of the $U(1)\times
U(1)\rtimes Z_2$ gauge theory in $4$ space-time dimensions.

It is amazing to see that the $d$-dimensional SPT phases and the
$d$-dimensional quantized topological terms in weakly coupled gauge theories are
classified by the same thing $\cH_B^{d}(G,\R/\Z)=H^{d+1}(BG,\Z)$.  In this
paper, we would like to clarify a duality relation between the $d$-dimensional
SPT phase and $d$-dimensional topological gauge theory for finite groups (see
section V).  (Here a topological gauge theory with a finite gauge group is
defined as a weakly coupled gauge theory with a finite gauge group and a quantized
topological term.) Such a  duality relation is exactly the duality relation
first proposed by Levin and Gu in 3-dimensions.\cite{LG1220}
We know that the
SPT phases are described by topological non-linear $\si$-models with symmetry
$G$:
\begin{align}
 \cL= \frac{1}{\la_s} [\prt g(x^\mu)]^2 + \imth W_\text{top}(g), \ \ g\in G
\end{align}
where the $2\pi$-quantized topological term $\int W_\text{top}(g)$ is
classified by $\cH_B^{d}(G,\R/\Z)$ which describes different SPT phases.
If we
``gauge'' the symmetry $G$, the topological non-linear $\si$-model will become
a gauge theory:
\begin{align}
 \cL= \frac{1}{\la_s} [(\prt -\imth A) g(x^\mu)]^2 + \imth W_\text{top}(g,A)
+\frac{(\t F_{\mu\nu})^2}{\la}.
\end{align}

If we integrate out $g$, we will get a pure gauge theory with a topological term
\begin{align}
 \cL= \frac{(\t F_{\mu\nu})^2}{\la} +\imth \t W_\text{top}(A)
.
\end{align}
This line of thinking suggests that the gauge-theory topological term
$\int \t W_\text{top}(A)$ is classified by
$\cH_B^{d}(G,\R/\Z)$. Since
$\cH_B^{d}(G,\R/\Z)=H^{d+1}(BG,\Z)$, the topological terms constructed this way agrees with the topological terms constructed through the
classifying space, which realizes the duality relation.

We remark that the
SPT phases are described by topological non-linear $\si$-models with symmetry
$G$.  If we ``gauge'' the symmetry $G$, the topological non-linear $\si$-model
will become a gauge theory-- a principal chiral model,\cite{LWchiral,CWchiral}
which realizes the duality relation.

\section{Lattice topological gauge theory} \label{ltgauge}

\subsection{Discretize space-time}
\label{disltgauge}

In this paper, we will consider gauge theories on discrete space-time which are
well defined.  We will discretize the space-time $M$ by considering its
triangulation $M_\text{tri}$ and define the $d$-dimensional topological
gauge theory on such a triangulation.  We will call such a theory a lattice
topological gauge theory.  We will call the triangulation $M_\text{tri}$ a
space-time complex, and a cell in the complex a simplex.

\begin{figure}[tb] \begin{center} \includegraphics[scale=0.6]{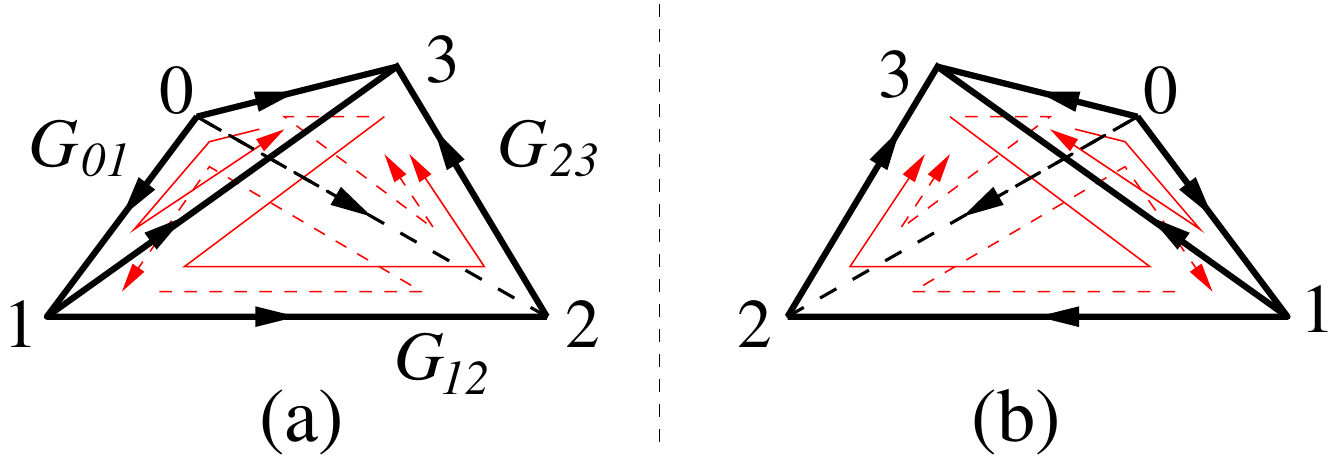} \end{center}
\caption{ (Color online) Two branched simplices with opposite orientations.
(a) A branched simplex with positive orientation and (b) a branched simplex
with negative orientation.  } \label{mir} \end{figure}

In order to define a generic lattice theory on the space-time complex
$M_\text{tri}$, it is important to give the  vertices of each simplex a local
order.  A nice local scheme to order  the  vertices is given by a branching
structure.\cite{C0527,CGL1172} A branching structure is a choice of orientation
of each edge in the $d$-dimensional complex so that there is no oriented loop
on any triangle (see Fig. \ref{mir}).

The branching structure induces a \emph{local order} of the vertices on each
simplex.  The first vertex of a simplex is the vertex with no incoming edges,
and the second vertex is the vertex with only one incoming edge, \etc.  So the
simplex in  Fig. \ref{mir}a has the following vertex ordering: $0,1,2,3$.

The branching structure also gives the simplex (and its sub simplexes) an
orientation.  Fig. \ref{mir} illustrates two $3$-simplices with opposite
orientations.  The red arrows indicate the orientations of the $2$-simplices
which are the subsimplices of the $3$-simplices.  The black arrows on the edges
indicate the orientations of the $1$-simplices.

\subsection{Lattice gauge theory on a branched space-time complex -- finite
gauge group}
\label{branchlattice}

Now, we are ready to define lattice gauge
theory on a branched space-time complex.  We first choose a gauge group $G$.
For the time being, let us assume that $G$ is finite.  We then assign
$G_{ij}\in G$ to each link $ij$ in the space-time complex $M_\text{tri}$, where
$i\to j$ is the orientation of the link.  For each simplex (with vertices
$i,j=0,1,...$), we assign a complex action amplitude $\al(\{G_{ij}\})$.

The imaginary-time path integral of the lattice gauge theory is given by
\begin{align} \label{ZGij}
Z={\sum_{\{G_{ij}\}}}' \prod_{[ij...k]}\al^{s(i,j,...,k)}(\{G_{ij}\}) ,
\end{align} where $\prod_{[ij...k]}$ is the product over all the simplices
$[ij...k]$ in the space-time complex $M_\text{tri}$, and $s(i,j,...,k)=1$ or
$\dag$ depending on the orientation of the simplex $[ij...k]$ defined by the
branching structure.  Two sets $\{G_{ij}\}$ and $\{G'_{ij}\}$ are said to be
gauge equivalent if there exists a set $g_i\in G$, such that \begin{align}
G'_{ij}=g_i G_{ij}g_j^{-1}.  \end{align} We require the above lattice theory to
be gauge invariant: \begin{align} \label{gaugeinv}
\prod_{[ij...k]}\al^{s(i,j,...,k)}(\{g_iG_{ij}g_j^{-1}\}) & =
\prod_{[ij...k]}\al^{s(i,j,...,k)}(\{G_{ij}\}), \nonumber\\ \forall g_i & \in
G, \end{align} for any \emph{closed} space-time complex $M_\text{tri}$:
$\partial M_\text{tri}=\emptyset$.  In \eqn{ZGij}, ${\sum}'_{\{G_{ij}\}}$ sums
over the gauge equivalent classes of $\{G_{ij}\}$.  This way, we define a
lattice gauge theory with gauge group $G$.  We may rewrite the path integral as
\begin{align} Z=|G|^{-N_v} \sum_{\{G_{ij}\}}
\prod_{[ij...k]}\al^{s(i,j,...,k)}(\{G_{ij}\}) ,
\label{pathI}\end{align} where $|G|$ is the
number of elements in $G$, $N_v$ the number of vertices.
The overall
normalization factor can be understood as modding out the theory by its overall
redundant phase volume.

\subsection{Lattice topological gauge theory -- finite gauge group}
\label{sec:finitegauge}

What is the low energy fixed-point theory of the
lattice gauge theory defined above, if the theory is in a gapped phase?  In a
study of SPT phases, we have discussed the gapped phases of non-linear
$\si$-model and the related fixed-point theories (or topological
theories).\cite{CGL1172,GW1248}  There, the fixed-point theories have the
following defining properties, that the action amplitudes for any paths are
always equal to 1 if the space-time manifold has a spherical topology.  Here we
will use the similar idea to study the fixed-point theory of the gapped phases
of a lattice gauge theory.

First, one possible low energy fixed-point theory is given by the following
action amplitude \begin{align} \label{VG1} \al(\{G_{ij}\})=1.  \end{align} Such
an  action amplitude does not change under the renormalization of the coarse
graining and describes a fixed-point theory.  However, such a fixed-point
theory describes a confined phase with trivial topological order.  In this
paper, we will regard such a fixed-point theory to have a trivial gauge group.

We would like to ask, is this the only way for a gauge theory to become gapped?
Can gauge theory become gapped without confinement and the reduction of gauge
group?  The answer to the above questions is yes:  a gauge theory with a finite
gauge group can be gapped even without confinement. So in this section, we will
consider gauge theories with a finite gauge group.

\begin{figure}[tb] \begin{center} \includegraphics[scale=0.5]{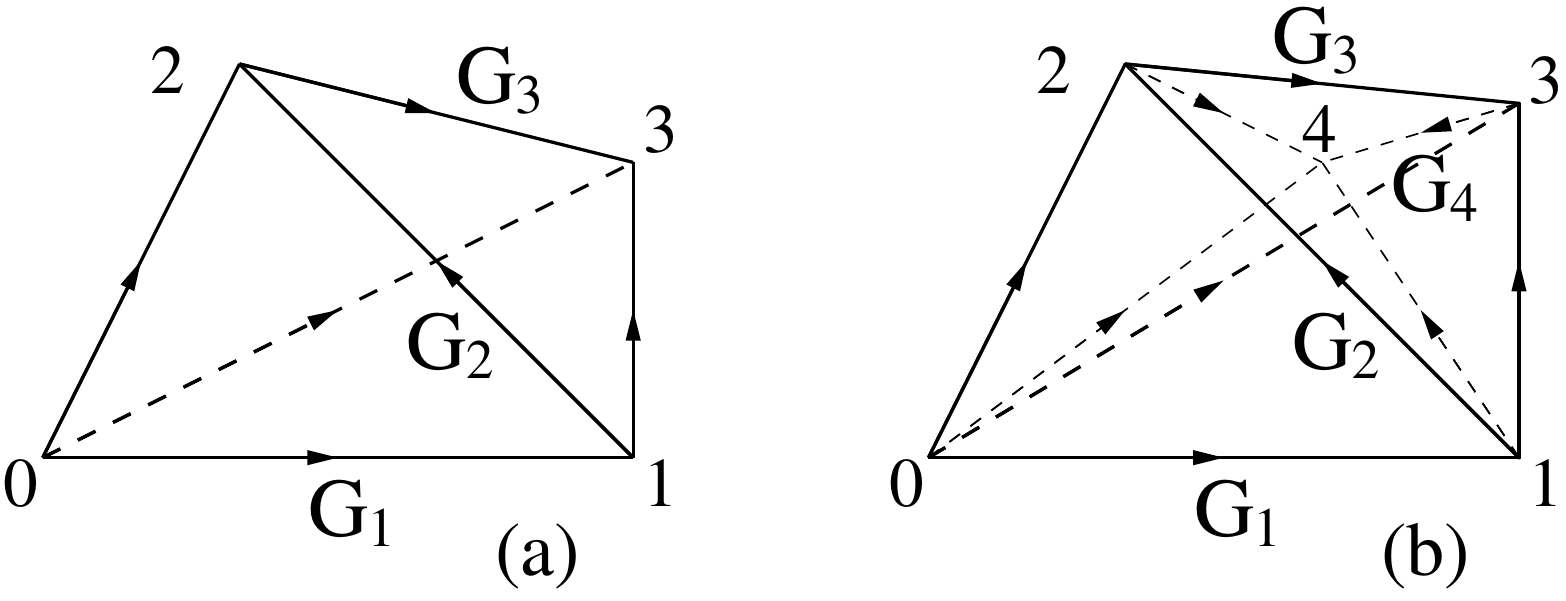}
\end{center}
\caption{ Complices with spherical topology: (a) a tetrahedron -- a (1+1)D
sphere, (b) a pentachoron -- a (2+1)D sphere.  The arrows on the link represent
the branching structure.  } \label{cocycnd} \end{figure}

The fixed-point action \eq{VG1} describes a confined phase since the Wilson
loop operator, such as $ G_{ij} G_{jk} G_{ki} $, can fluctuate strongly.  So to
obtain a fixed-point theory with the original gauge group $G$, we require that
\begin{align}
|\al(\{G_{ij}\})|=1\ \text{ only when }  G_{ij} G_{jk} G_{ki}=1,
\end{align}
on all the triangles of the simplex on which $\al(\{G_{ij}\})$ is defined.
In other words, the amplitude of a path is zero
if there is a non-zero flux $ G_{ij} G_{jk} G_{ki} \neq 1 $
on some trangles.
We will call this condition a flat connection condition since it
corresponds to requiring the ``field strength $F=0$''.  In order for
$\al(\{G_{ij}\})$ to describe a fixed-point topological theory, we also
require that \begin{align} \label{cocy}
\prod_{[ij...k]}\al^{s(i,j,...,k)}(\{G_{ij}\})=1 \end{align} on all the complex
$M_\text{tri}$ that have a spherical topology.  (It would be too strong to
require $ \prod_{[ij...k]}\al^{s(i,j,...,k)}(\{G_{ij}\})=1$ on any closed
complex $M_\text{tri}$.)

In (1+1) dimensions, the simplest sphere is a tetrahedron.  Due to the flat
connection condition, we can use $G_1=G_{01}$, $G_2=G_{12}$, and $G_3=G_{23}$
to label all the $G_{ij}$'s (see Fig. \ref{cocycnd}a).  For example,
$G_{02}=G_1G_2$.  On a tetrahedron, the condition \eq{cocy} becomes
\begin{align} \frac{ \al(G_1,G_2) \al(G_1G_2,G_3) }{ \al(G_2,G_3)
\al(G_1,G_2G_3) }=1.  \end{align} We note that if $\al(G_1,G_2)$ is a solution
of the above equation, then $\al'(G_1,G_2)$ defined below is also a solution of
the above equation: \begin{align} \label{3cobound} \al'(G_1,G_2) =
\al(G_1,G_2) \frac{\bt(G_1)\bt(G_2)}{\bt(G_1G_2)}.  \end{align} We regard
the two solutions to be equivalent.  The equivalent classes of the solutions
correspond to different lattice topological gauge theories in (1+1) dimensions.

In (2+1) dimensions, the simplest 3-sphere is a pentachoron.  Due to the flat
connection condition, we can use $G_1=G_{01}$, $G_2=G_{12}$, $G_3=G_{23}$, and
$G_4=G_{34}$ to label all the $G_{ij}$'s (see Fig. \ref{cocycnd}b).  For
example, $G_{02}=G_1G_2$.  On a tetrahedron, the condition \eq{cocy} becomes
\begin{align}  \label{pentagon} \frac{ \al(G_1,G_2,G_3) \al(G_1,G_2G_3,G_4)
\al(G_2,G_3,G_4) }{ \al(G_1G_2,G_3,G_4) \al(G_1,G_2,G_3G_4) }=1 \end{align} We
note that if $\al(G_1,G_2,G_3)$ is a solution of the above equation, then
$\al'(G_1,G_2,G_3)$ defined below is also a solution of the above equation:
\begin{align} \al'(G_1,G_2,G_3) = \al(G_1,G_2,G_3) \frac{ \bt(G_1,G_2) \bt(G_1G_2,G_3)
}{ \bt(G_2,G_3) \bt(G_1,G_2G_3) } .  \end{align} Again, the above defines an
equivalence relation between the solutions.  The distinct equivalence classes of the
solutions correspond to different lattice topological gauge theories in (2+1)
dimensions.

The above discussion can be generalized to any dimensions.  We also note that
the equivalence class defined above is nothing but the group cohomology class
$\cH^{d}(G,\R/\Z)$. Therefore, the lattice topological gauge theory with a
finite gauge group $G$ is classified by $\cH^{d}(G,\R/\Z)$ in $d$
dimensions, a result first obtained by Dijkgraaf and Witten.\cite{DW9093} From
the above discussions, we see that the result can also be phrase in a more
physical way: the gapped phases of a lattice gauge theory with a finite gauge
group $G$ is classified by $\cH^{d}(G,\R/\Z)$ in $d$ space-time dimensions, provided
that there is no confinement and the reduction of gauge group (\ie the ``field
strength $F$'' fluctuate weakly).

We see that a lattice gauge theory can have many different gapped phases.  One
kind of gapped phases have no confinement nor reduction of gauge group (say due
to the Higgs mechanism).  This kind of  gapped phases are classified by
$\cH^{d}(G,\R/\Z)$ for finite group and in $d$ space-time dimensions.
Other kind of gapped phases may have confinement or reduction of gauge group.
Those gapped phases may be described by $\cH^{d}(G',\R/\Z)$ where $G' \subset
G$ is the unbroken gauge group.

\section{Topological gauge theory as a non-linear $\si$-model with classifying
space as the target space} \label{top_sigmamodel}

\subsection{Classification of
$G$-bundles on a $d$-manifold $M$ via classifying space $BG$  and universal
bundles $EG$ of group $G$}

In order to define topological gauge theory for continuous group (as well as
for finite group), Dijkgraaf and Witten pointed out that all the gauge
configurations on $M$ can be understood through classifying space $BG$  and
universal bundles $EG$ (with a connection): all $G$-bundles on $M$ with all the
possible connections can be obtained by choosing a suitable map of $M$ into
$BG$, $\gamma: M \to BG$.\cite{DW9093} $BG$ is a very large space, often
infinite dimensional.
If we pick a connection in the
universal bundle $EG$, even the different connections in the same $G$-bundle
on $M$ can  be obtained by different maps $\gamma$.  Therefore, we can express
the imaginary-time path integral of a gauge theory as
\begin{align}
Z=\sum_\ga \e^{\imth S[\ga]}
\end{align}
where $\sum_\ga$ sum over all the maps $\ga$: $M\to BG$,
and $S[\ga]$ is the action for the map $\ga$.  The dynamics of the gauge theory
is controlled by the action $S[\ga]$ and the connection on $BG$.  In other
words, once we specify a connection on $BG$, every map $\ga: M\to BG$ will
define a connection $F$ on $M$. Thus we can view the action $S[\ga]$ as a
function of the connection, $S[F]$ (plus, possibly other gauge invariant
degrees of freedom).  We see that, in some sense, a gauge theory can be viewed
as a non-linear $\si$-model with classifying space $BG$ as the target space.

We like to remark that when we study gauge theory in a fixed space-time
dimension $d$, we can choose a truncated classifying space $BG^n$ which has a
finite dimension and a finite volume.  We can view  a gauge theory as a
non-linear $\si$-model with the truncated classifying space $BG^n$ as the
target space.

In the following, we will use this point of view to study topological gauge
theory.  We have to say that such an approach is quite indirect compared to the
discussion in section \ref{ltgauge}.  But, as we will see later, the two
approaches give rise to the same classification of topological gauge theories
for finite gauge groups.

\subsection{Topological gauge theory from the
non-linear $\si$-model of $BG$}

Viewing a gauge theory as a non-linear $\si$-model with classifying space $BG$
as the target space, we can study topological terms in the gauge theory by
studying the topological terms in the corresponding non-linear $\si$-model.
Here, we write $S[\ga]$ as $S[\ga]=S_\text{top}[\ga]+\imth S_\text{dyn}[\ga]$.  The
term $S_\text{top}[\ga]$ is independent of space-time metrics and is called the
topological term.  We are mainly concerned about the question whether the systems
described by $S[\ga]=S_\text{top}[\ga]+\imth S_\text{dyn}[\ga]$ and
$S_0[\ga]=\imth S_\text{dyn}[\ga]$ are in the same phase or not.
In general, a quantized topological term $=S_\text{top}[\ga]$ may make
$S[\ga]$ and $S_0[\ga]$ to describe different phases.
So we may gain some understanding of quantum phases
by studying  quantized topological terms.

In \Rf{CLW1141,CGL1172}, we studied the $2\pi$ quantized topological
$\th$-terms in lattice non-linear $\si$-model with the symmetry group $G$ as
the target space.  We find that such  quantized topological terms are
classified by Borel cohomology classes $\cH_B^{d}(G, \R/\Z)$.  In this case,
the different quantized topological terms do give rise to different quantum
phases.  Here, we can use a similar approach to construct/classify topological
terms in non-linear $\si$-model with classifying space $BG$ as the target
space.

To use the above idea to study lattice gauge theories,
we need to put the above discussion on a
lattice by trianglating the space-time manifold $M$ into a complex
$M_\text{tri}$.  The mapping $\ga$ from $M$ to $BG$ now becomes a mapping from
$M_\text{tri}$ to  $BG$.  However, the  mapping from $M_\text{tri}$ to  $BG$
can be defined differently, with extra structures and information in some
definitions as oppose to others.

We may define the map  from $M_\text{tri}$ to  $BG$ as a map from the vertices
of $M_\text{tri}$ to  $BG$.  We have chosen such kind of map when we use
lattice topological non-linear $\si$-model with the symmetry group $G$ as the
target space to classify the SPT phases.  However, such maps are not adequate
to define lattice gauge theory, since the maps of the vertices do not allow us
to obtain a connection on $M_\text{tri}$ by pulling back the connection on
$BG$.

To define a lattice gauge theory where gauge degrees of freedom reside
on the edges of the triangulation $M_\text{tri}$,
the map therefore need at least to specify how the set of 1-simplices,
in $M_\text{tri}$ is mapped into $BG$. In principle, no further
detail is necessary to define the gauge theory. However, we will
take a less general route and instead regard the map $\ga$ as an
\emph{embedding} of $M_\text{tri}$ into $BG$. This means that information
about the mapping of all the higher simplices, such as 2d faces that connect
the edges are also completely specified.
As will be evident in more detailed discussion of the Mathematics of the construction
in section \ref{connectI_II}, such a choice of map requires that the lattice gauge
theory is in the semiclassical limit where the fluctuations in the field
strength are weak. 
In this case, the connection on $BG$ naturally
becomes a connection on $M_\text{tri}$.  Different embeddings correspond to
different gauge field configurations on $M_\text{tri}$.  In order to write down
an action $S_\text{top}$ for the lattice topological gauge theory on a $d$-dimensional
complex $M_\text{tri}$, we assign a $U(1)$ phase $\th_i$ mod $2\pi$ to each
$d$-dimensional simplex in the triangulated classifying space $BG$.  Such
an assignment correspond to a $d$-cochain $\mu_{d}$ in
$C^{d}(BG,\R/\Z)$.  Then, the  action  $S_\text{top}$ is the sum of the $U(1)$ phases
$\th_i$ on the simplices in $\ga(M_\text{tri}) \in B$.  The resulting
total phase $S_\text{top}$ corresponds to evaluating the cochain $\mu_{d}$ on the complex
$\ga(M_\text{tri})$:
\begin{equation}
\label{lattgauge}
S_\text{top}=2\pi \< \mu_{d}, \ga(M_\text{tri})\> \ \ \text{mod } 2\pi.
\end{equation} Such an  action
amplitude  $\e^{\imth S_\text{top}[\ga]}$ depends on the embedding $\ga$ and defines a
dynamical gauge theory.  This way, we write a lattice gauge theory as a lattice
non-linear $\si$-model with $BG$ as target space, through the embedding map
$\ga$.

To define a lattice topological term, we may choose $S_\text{top}[\ga]=0$
mod $2\pi$ for any maps $\ga$ as long as $M_\text{tri}$ has no boundary.  This
is the action that we choose to classify the SPT phase using lattice
topological non-linear $\si$-model.

But here, we like to choose a more general topological term
$S_\text{top}[\ga]$.  As a topological term, $S_\text{top}[\ga]$ should not
depend on the ``metrics'' of the complex $M_\text{tri}$ (\ie the size and the
shape of the $M_\text{tri}$).  We would also like to consider restricting
$S_\text{top}[\ga]$ such that it has no dependence on the connection on
$M_\text{tri}$, as long as $M_\text{tri}$ has no boundary.  But
$S_\text{top}[\ga]$ may depend on the topology of $M_\text{tri}$, or more
precisely on the homological class of the embedding $\ga(M_\text{tri})$ in
$BG$.  Those considerations suggest that we can define a topological action by
choosing a cocycle $\al_{d} \in Z^{d}(BG,\R/\Z)$:
\begin{equation}
\label{topgauge}
S_\text{top}[\ga]=2\pi \< \al_{d}, \ga(M_\text{tri})\> \ \ \text{mod } 2\pi.
\end{equation}
Note that the  $d$-cocycle are special $d$-cochains
whose evaluation on any  $d$-cycles [\ie $d$-dimensional closed
complexes] are equal to $0$ mod 1 if the $d$-cycles are boundaries of some
$(d+1)$-dimensional complex.  So, each $d$-cocycle $\al_{d}$ in
$Z^{d}(BG,\R/\Z)$ defines a lattice topological gauge theory in
$d$-dimensions.

If two $d$-cocycles, $\al_{d}, \al'_{d} \in Z^{d}(BG,\R/\Z)$, differ by a
coboundary: $\al'_{d} - \al_{d} = \dd \mu_d$, $\mu_d \in C^d(BG,\R/\Z)$, then,
the corresponding action amplitudes, $\e^{\imth S_\text{top}[\ga]}$ and $\e^{\imth
S_\text{top}'[\ga]}$, can smoothly deform into each other without phase transition.
So
$\e^{\imth S_\text{top}[\ga]}$ and $\e^{\imth S_\text{top}'[\ga]}$, or $\al_{d}$ and $\al'_{d}$,
describe the same quantum phase.  Therefore, we regard $\al_{d}$ and $\al'_{d}$
to be equivalent.  The equivalent classes of the $d$-cocycles form the $d$
cohomology class $H^{d}(BG,\R/\Z)$.  We conclude that the  topological terms in
weakly coupled lattice gauge theories are described by  $H^{d}(BG,\R/\Z)$
in $d$ space-time dimensions.

For finite gauge group, we can choose a flat connection for the $G$-bundle
$EG$.
Given that, the connection on
$M_\text{tri}$ is always flat regardless of the embedding $\ga$.
In this case, the
topological gauge theory defined via the classifying space $BG$ is closely
related to the lattice topological gauge theory defined in section
\ref{ltgauge}.  On the other hand, we can also choose a non-flat  connection
for the $G$-bundle $EG$.  In this case, the different  embeding $\ga$ will give
rise to different  connections on $M_\text{tri}$.  So the gapped phases of the
gauge theory classified by  $H^{d}(BG,\R/\Z)$ can appear even when there are
weak fluctuations of the ``field strength $F$''.  Certainly, those  gapped
phases can also  appear when the ``field strength $F$'' are zero, as discussed
in section \ref{ltgauge}.
For finite  group $G$, we have $H^{d}(BG,\R/\Z)\simeq H^{d+1}(BG,\Z)\simeq
\cH_B^d(G,\R/\Z)$ (see \eqn{HdHd1}).

For continuous gauge group, the connection for the  $G$-bundle $EG$ is always
non-flat.  In this case, the different  embeddings $\ga$ always give rise to
different  connections on $M_\text{tri}$.  So the the gauge theory in general
contain fluctuations of the ``field strength $F$''.

In appendix \ref{HBGRZ}, we show that
$H^d (BG,\R/\Z)$ has a form $H^d (BG,\R/\Z)=\R/\Z\oplus ... \oplus
\R/\Z \oplus Z_{n_1} \oplus Z_{n_2}\oplus ...  $.
So for continuous groups, $H^d (BG,\R/\Z)$ may not be discrete
and the corresponding topological terms are also not quantized.
So the quantized topological terms are described by the discrete part of
$H^d (BG,\R/\Z)$:
\begin{align}
 \text{Dis}[H^d (BG,\R/\Z)]&=Z_{n_1} \oplus Z_{n_2}\oplus ...
\nonumber\\
&= \text{Tor}[H^{d+1}(BG,\Z)]
\end{align}
(see \eqn{DisH}).  (Note that for finite group $\text{Dis}[H^d (BG,\R/\Z)]=H^d
(BG,\R/\Z)=\text{Tor}[H^{d+1}(BG,\Z)]$.) We can use the torsion of the cohomology class
$\text{Tor}[H^{d+1}(BG,\Z)]$ of the classifying space $BG$ to construct the quantized
topological terms.

\subsection{The relation between the first and the second constructions}

For finite gauge group $G$, its classifying space has a property $\pi_1(BG,\Z)
\simeq G$.  So, each non-trivial loop in $BG$ can be associated with a
non-trivial element in $G$, while the trivial loop (or a point) is associated
with the identity element in $G$.  For continuous group, we can choose a
one-to-one mapping between the non-trivial elements in $G$ and a set of loops
in $BG$ that all go through the  base point in $BG$.  As an element approaches
the identity, its loop shrinks to the base point.  Using such a property, we
can understand the relation between the first and the second constructions
discussed above.

The lattice gauge theory in the first construction is defined on a space-time
complex $M_\text{tri}$.  A lattice gauge configuration is given by a set of
group elements, $\{G_{ij} \in G\}$, on each link $ij$.  So a lattice gauge
configuration corresponds to a 1-skeleton in $BG$.  The 1-skeleton is formed by
the loops that correspond to $G_{ij}$.

A triangle $(ijk)$ in $M_\text{tri}$ is mapped to a loop in $BG$ using the
above correspondence. If the gauge configuration is flat:
$G_{ij}G_{jk}G_{ki}=1$, the loop is contractible.  If $G$ is finite
$\pi_n(BG)=0$ for $n>1$.  So there is a unique way to extend the above
contractible loop to a disk in $BG$. This way, we extend the 1-skeleton to a
2-skeleton. Since $\pi_3(BG)=0$, we can extend the  2-skeleton to 3-skeleton,
\etc.  Therefore, for a finite group, we can obtain a canonical map from  a
lattice gauge configuration $\{G_{ij}\}$ to an embedding map $\ga: M_\text{tri} \to
BG$.  Such an embedding map relate the group cohomology cocycle
$\al(G_1,...,G_d)$ for the group $G$ to the topological cocycle $\al_{d}$ in
$BG$.  So there is a clear one-to-one relation between the first and the second
construction for finite gauge groups.

For continuous groups, $\pi_n(BG)$ are non-trivial. So the relation between the
second construction and lattice gauge theory is less clear.  For a lattice
gauge configuration with $G_{ij} \approx 1$, there is a unique way to extend
the  1-skeleton to an embedding map $\ga: M_\text{tri} \to BG$.  For example,
even when $\pi_2(BG)\neq 0$, we can still uniquely extend a small triangle to a
disk with the smallest area.

We can use this idea to find a map from  a lattice gauge configuration $G_{ij}$
to an embedding map $\ga$ by choosing the extension with the minimal
area/volume.  The topological action
$S_\text{top}[\ga]=S_\text{top}[\{G_{ij}\}]$ obtained this way is topological
at least when $G_{ij} \approx 1$.  We can extend $S_\text{top}[\{G_{ij}\}]$ to
any values of $G_{ij}$ far from $1$ and still
keep its topological properties. The resulting
$S_\text{top}[\{G_{ij}\}]$ may not be a continuous function of the  lattice
gauge configuration $G_{ij}$.  But it is a measurable function (\ie the
discontinuity happens only on a measure-zero set).

\section{Differential character and topological gauge theory in $d$ = odd
space-time
dimensions}

In the last section, we constructed topological terms in
a weakly coupled gauge theory assuming that
the action $S$  does not depend on the connection on the space-time complex
$M_\text{tri}$, as long as $M_\text{tri}$ has no boundary.  In this section we
are going to relax such a restriction and allow the action to  depend on the
gauge connection for $d$ = odd space-time dimensions.  However, we will still
assume that the action is independent of the ``metrics'' of  $M_\text{tri}$,
which ensure the constructed term to be topological.  Such a
generalized topological term corresponds to a Chern-Simons
term.\cite{DW9093}
For simplicity, the the rest of this section, we will concentrate
on $d$ = 3 space-time dimensions. However, the results and approaches
can be easily generalized to any odd dimensions.

\subsection{3$d$ Chern-Simons theory}
\label{CSF}

First, let us define the  Chern-Simons theory carefully.  Naively, a
Chern-Simons theory of gauge group $G$ on a closed 3$d$ space-time manifold
$M$ is defined by the action
\begin{equation}
S_{CS}=\int_M \frac{K}{4\pi} \Tr(AF-\frac{1}{3}A^3) .
\end{equation}
However, such a definition is
incomplete, since for some smooth gauge configurations $F$, the gauge potential
$A$ cannot be well defined smooth functions on $M$.  To fix this problem, we
may try to view $M$ as the boundary of $B$:  $\prt B=M$, and try to define the
Chern-Simons theory action as\cite{DW9093}
\begin{align}
S_{CS}=\int_B \frac{K}{4\pi} \Tr(F^2) .
\end{align}
But it may not be always possible to
extend the gauge configuration on $M$ to $B$.  Let us assume that the boundary
of $B$ is $n$ copies of $M$: $\prt B=M^n$, and let us assume that for a proper
$n$, the gauge configuration on $M^n$ can be extended to $B$.  In this case, we
can define the  Chern-Simons theory action as\cite{DW9093}
\begin{align}
\label{CSFn}
S_{CS}=\frac{1}{n} \int_B \frac{K}{4\pi} \Tr(F^2) .
\end{align}
In the following, we will implement the above idea  more rigorously, which
allow us to define a generalized  Chern-Simons in any $d=$ odd space-time
dimensions and for any gauge group $G$.

\subsection{3$d$ Chern-Simons theory of gauge group $G$}

In our brief discussion of constructing Chern-Simons terms above, we have introduced
the need for a four dimensional manifold $B$ in which $M$ embeds. A most natural
choice, given our task to classify these terms that depends on
gauge connections, would be to choose some $B$
inside the classifying space $BG$, such that $\prt B=\ga(M^n)$.

To understand how the integer $n$ emerges, let us consider
$3$-homology class $H_{3}(BG,\mathbb{Z})$ of the classifying
space $BG$.  This classifies the obstruction for a given closed three
manifold to be the boundary of some four manifold in $BG$.
For a finite group $G$ however, $H_d(BG,\mathbb{Z})$ contains only
torsion.\footnote{A torsion element $X$ of order $n$ is one such that $n\cdot X
=0$.} For continuous group, $H_{d}(BG,\mathbb{Z})$ also contains only torsion
if $d$ is odd.  Thus $H_{3}(BG,\mathbb{Z})$ contains only torsion.

Let $M$ be 3-dimensional and let $n$ be the integer such that $n \cdot
H_3(BG,\mathbb{Z})=0$.  So for any embedding $\ga(M)$, $[\ga(M)]^n$ is a
boundary of $4$-dimensional complex $B$: $\prt B=  [\ga(M)]^n$ inside $BG$.
Following the idea in section \ref{CSF}, a suitable action $S$ of the
Chern-Simons theory
is given by\cite{DW9093}
\be
\label{action}
\frac{S_{CS}}{2\pi} = -\frac{1}{n} \left(\int_B \frac{K}{4\pi^2}F^2 -
\langle \om \,,B \rangle \right)\,\, \textrm{mod } 1 ,
\ee
for some  $\om \in
H^{4}(BG, \mathbb{Z})$. This definition works both for finite and continuous
compact groups.  One can see that Eqn. \eq{action} is basically the
Chern-Simons action \eq{CSFn}. We note that the choice of the pair $K$ and $\om$
defines the theory. However, they are not independent.
In fact, $K$ has to be chosen such that
$\int_B \, \frac{K}{4\pi^2} F^2 - \langle \omega \,,B
\rangle =0 $ for \emph{all} closed manifolds $B$.\footnote{Mathematically, we are picking
out the image of $\omega$ in $H_4(BG,\mathbb{R})$ via the Weil homomorphism.}
This implies that the action is in
fact exact, and the theory is truly three dimensional, which contrasts with WZW theories.
In other words, there must
be some analogue of Chern-Simons forms $\theta(A)$ depending on the connection
$A$,\footnote{One Mathematical detail that should be noted here is that the
connection $A$ on $BG$ has a canonical choice, called the \emph{universal
connection} $A_u$. Choosing $A=A_u$, all possible connections on $M$ can be
obtained by picking a corresponding embedding $\ga$. Therefore on $M$, the
connection can be viewed simply as a function of $\ga$. } such that
\be
\label{diffchar}
\dd \theta(A) = \frac{K}{4\pi^2}F^2 - \omega,
\ee
and that the action can be rewritten as
\be
\frac{S_{CS}}{2\pi} = \langle \tilde{\theta}(A), \ga(M) \rangle,
\ee
where $\tilde{\theta}= \theta\,\,\textrm{mod } 1$.
The connection
evaluated on $\ga(M)$ is determined by the embedding $\ga$. Therefore $S_{CS}$
is a function of $\ga$. i.e. We write $S_{CS} \equiv S_{CS}[\ga]$.  It turns
out that indeed $\theta(A)$ exists, and the corresponding
$\tilde{\theta}(A)$, is called the differential characters, which
is uniquely determined for given $\omega$ for compact groups.

Therefore
the Chern-Simons action is classified by $H^{4}(BG,\mathbb{Z})$.  Having
defined the action, the path-integral is given by a sum over embedding $\ga$,
corresponding to a sum over different bundles and connections on $M$
\be Z =
\sum_{\ga} e^{\imth S_{CS}[\ga]-S_\text{dyn}}.
\ee
We can see that in this formulation of
the Chern-Simons theory, its connection with the non-linear sigma model
discussed in section \ref{top_sigmamodel} is very explicit, where space-time
manifold $M$ is embedded in the target space $BG$ with the embedding $\gamma$.
Eqn. (\ref{action}) however, is sensitive to the connection, therefore relaxing
the requirement in section \ref{top_sigmamodel}.
Consider several limiting cases.
For simply connected compact groups, such as $SU(2)$, there is no non-trivial
torsion, and $n=1$. The term involving $\om$ in Eqn. \eq{action} contributes
only to an integer and thus becomes trivial, and the action exactly reduces to
\eq{CSFn}. i.e.  The differential character reduces to the Chern-Simons form.
On the other hand, by
comparing with Eqn. (\ref{topgauge}), we realize that when $F=0$, the
differential character $\tilde{\theta}(A)$ reduces to a cocycle $\alpha$ in
$H^{d}(BG, \mathbb{R}/\mathbb{Z})$, and thus coincide with the non-linear
sigma model.  In other words, the non-linear sigma model in $3d$
forms only a subset of
the Chern-Simons theory. In the case of a finite group however
$F \in H^d(BG,\mathbb{R})\equiv 0$, and $H^{3}(BG,
\mathbb{R}/\mathbb{Z})$ is isomorphic to $H^4(BG,\mathbb{Z})$. Thus
in these cases the non-linear sigma models is in one-to-one correspondence
with the Chern-Simons theories.
As we will
discuss in the next subsection, this is in fact precisely the topological
lattice gauge theory.

\subsection{3$d$ topological lattice gauge theory and Chern-Simons theory}
\label{connectI_II}

In this section, we would like to make connection between the Chern-Simons
theory for finite group $G$ defined in the previous section and the topological
gauge theory in section \ref{ltgauge}. The discussion here closely parallels
that in \Rf{DW9093}.

In the remaining part of this paper, we will only consider the case of finite
gauge groups.  In the case of finite groups, we can choose $S_\text{dyn}$ such
that there is no non-trivial field strength, by setting $ S_\text{dyn} \to
\infty$ for any configurations with  non-trivial field strength.  In this case,
finite field strength gives rise to gapped excitations.  So the low energy
physics below the gap is controlled by $F=0$ configurations.  We choose
$S_\text{dyn}=0$ for  configurations with zero field strength.  Since in the
following, we will limit ourselves to  zero-field-strength configurations only,
we will drop $S_\text{dyn}$.

Those $F=0$ field configurations can be characterized by
Wilson loops, corresponding to maps from the fundamental group $\pi_1(M)$ to
$G$.  This assignment of group element on each loop in $M$ depends on which
loop in $BG$ it is mapped to. In other words, the assignment depends entirely
on the embedding $\ga$, since each homotopy class of loops in $BG$ is assigned
a unique element $g \in G$.\footnote{This follows from the property of $BG$
that $\pi_1(BG)$ is isomorphic to $G$. Homotopically equivalent $\ga$ therefore
give rise to the same assignment of group elements. }
Also, as already noted in the previous section, in this case the differential
character also reduces to a 3-cocycle $\alpha$ in $H^3(BG, U(1))$. The
path-integral can then be understood as \begin{eqnarray}\label{CSfinite} Z &=&
\sum_{\{g_i\}, i\in \pi_1(M)} e^{\imth S_{CS}[\{g_i\}]} \nonumber \\ &=&
\sum_{\{g_i\}, i\in \pi_1(M)} \langle \ga^*\alpha(\{g_i\}), M\rangle,
\end{eqnarray} where $\{g_i\}$ is the set of group elements in $G$ assigned to
each homotopy class of loops in $M$, and we have rewritten the dependence of
the Chern-Simons action on the embedding as a dependence on the set $\{g_i\}$.
An admissible set $\{g_i\}$ is not arbitrary, as we will explain in more detail
in some simple examples later. In fact, they form a representation of the
fundamental group $\pi_1(M)$.  This action is already very suggestive that we
are dealing with a topological lattice gauge theory. To make precise the
connection with the lattice theory, one needs to triangulate the space-time
manifold $M = \sum_i \epsilon_i T_i$ for $d$ simplices $T_i$ each with some
orientation $\epsilon_i$, and  demonstrate that Eqn. (\ref{CSfinite}) can be
broken down into local contributions from each simplex. This can indeed be
achieved in two steps.  \subsubsection{Path-Integral on a single simplex}
First, one needs to define the path-integral for a single simplex $T_i$.  A
path-integral for a simplex is one for which the manifold concerned has
boundaries.  Let us comment briefly on the physical meaning of a path integral
on a manifold with a boundary.  Consider a $d$ space-time manifold $M$ with a
single boundary $\Sigma_{d-1}$, on which we need to specify boundary conditions.
i.e. we fix the boundary value of the embedding map $\gamma(\Sigma)$, and we
only sum over all maps which reduce to the boundary value in the path
``integral''.  The boundary has thus led to some physical degrees of freedom
that reside at the $d-1$ boundary, to which one can associate with it a Hilbert
space $\mathcal{H}(\Sigma_{d-1})$ and the path integral $Z$ with specific boundary
condition can then be understood as the wavefunction that describes a
particular state defined on a given ${d-1}$ dimensional fixed-time slice. Here we
are identifying the direction orthogonal to $\Sigma$ as time.  Note that if $M$
has multiple boundaries, a Hilbert space would be associated to each boundary,
and the path-integral is a multi-linear map that maps $\prod_i
\mathcal{H}_{\Sigma^i_{d-1}}$ to a phase.  Let us now return
to the path-integral of a
single simplex. For concreteness, consider $d=3$ and the $d$ simplex is a
tetrahedron. The surface of a tetrahedron is bounded by four triangles with six
edges connecting four vertices.  This provides a convenient way to obtain a
basis for the Hilbert space on the surface of the tetrahedron. The idea is that
a base point $x$ is chosen in $BG$, such that for each embedding $\ga$ they are
deformed to make sure that all the vertices in the tetrahedron (and ultimately
the entire triangulation of space-time $M$) are mapped to $x$.  Every edge
$e_{ab}$ connecting vertices $a,b$ is then mapped to a loop in $BG$, and as
discussed earlier, each edge can be assigned a group element $g_{ab}$.  In
practice, to specify the state uniquely we also need to give an orientation to
the edge.  The same state denoted $g_{ab}$ with a given orientation can be
equally represented by $g_{ab}^{-1}$ but whose orientation is reversed. One way
to fix the orientation is to number the vertices, such that the arrow attached
to each edge points toward the vertex taking the larger index, and we uniquely
label the element as $g_{ab}$ for $a<b$.  This is precisely the branching
structure already discussed in section \ref{branchlattice}.  The three edges
binding a triangle do not form a closed loop, and that the edges between $ab$
and $ac$ for $a<b<c$ determine an orientation for the triangle.  The
orientation of each tetrahedron $\epsilon$, with vertices $v_{a,b,c,d}$ for
$a<b<c<d$, can be identified with $s(a,b,c,d)$ defined there (understanding $s=
\dag$ as $\epsilon = -1$).

Consider one of the triangles on the tetrahedron bounded by three vertices
$v_a, v_b, v_c, a<b<c$.  The three elements $g_{ab},g_{bc},g_{ac}$ attached to
the three edges of the triangle is subjected however to the condition that \be
\label{vanishflux} g_{ab} g_{bc} (g_{ac})^{-1} = 1.  \ee Here $g_{ac}$ is
inverted because its orientation is opposite to the orientation defined by the
edges $ab$ and $bc$.  Note that for general non-Abelian groups $G$, the order
of multiplication follows the arrows of the edges.  The above relation follows
from the fact that the triangle is mapped to a 2 manifold in $BG$ which is
topologically a 2-sphere with a marked point $x$.  Moving along the three edges
following the orientations however leads to a contractible loop on the 2-sphere
which should be associated to the identity.  The construction automatically
reproduces the flatness requirement of the topological lattice gauge theory.
The constraints mean that of the six edge elements on the tetrahedron, only
three are independent.  The path-integral on the tetrahedron is a $U(1)$ phase
that depends on the surface state given by the set of group elements
$\{g_{ab}\}$ attached to the edges. It is given by  the 3-cocycle
$\alpha^{s(a,b,c,d)}(\{g_{ab}\}) = \langle \alpha, \gamma(T_i)\rangle$
evaluated on the image of the tetrahedron in $BG$. Again we make explicit its
dependence on the field configuration $\{g_{ab}\}$ through the embedding $\ga$.
Recall that only three elements are independent, let us also write
$\alpha(\{g_{ab}\}) = \alpha(g_{ab},g_{bc},g_{cd})$, for $a<b<c<d$.

\subsubsection{Gluing relations} Having defined the path-integral on a single
simplex, we need to glue them together.  Path-integrals defined on manifolds
with boundaries, satisfy the so called gluing relations \be Z_{M_c} = \sum_i
Z_{M_a}(v_i) Z_{M_b}(v^i) \ee where $M_a$ and $M_b$ has a common boundary
$\Sigma$, and $v_i$ are the basis vectors of states on $H_\Sigma$.  The
manifold $M_c$ is then the manifold formed from gluing together $M_a$ and $M_b$
along $\Sigma$. Here we should be careful with orientations, and strictly
speaking the gluing along $\Sigma$ is such that it is \emph{out-going} on
$M_a$, and \emph{in-going} (with reversed orientation) in $M_b$. In other
words, it means the full path integral is given by taking the products of path
integrals over the submanifolds sharing boundaries, and summing over states in
the Hilbert spaces defined on the shared boundaries.  Therefore, we finally
have
\be \label{gluing}
Z[M] = |G|^{-N_{v}}\sum_{\{g_{ab}\}} \prod_i \alpha^{s_i}(\{g_{c
d}\}, \{v_{c,d} \in T_i\}).
\ee
where again $N_{v}$ is the total number of vertices.  Since $\alpha$ is a
3-cocycle, the path-integral should simply give 1 on a 3-sphere, which is the
boundary of a 4-ball.  This is precisely the same consideration discussed
already in section \ref{sec:finitegauge}. This means $\alpha$ satisfies Eqn.
(\ref{pentagon}), which is as expected since $H^3(BG,\R/\Z) \cong
\cH_B^3(G,\R/\Z)$. The normalization, together with the pentagon relations Eqn.
(\ref{pentagon}) ensures that there is no dependence on the choice of
triangulation of $M$.  This path-integral enjoys gauge invariance as in Eqn.
(\ref{gaugeinv}), and that the rescaling symmetry in Eqn. (\ref{3cobound})
follows from the fact that $\langle \alpha, B \rangle$ is invariant under
$\alpha \to \alpha\delta\beta$ on a closed manifold $B$.  This completes the
connection between the Chern-Simons theory and the topological lattice gauge
theory for finite group $G$.

\section{Duality relation between topological lattice gauge theory and SPT
phases for finite gauge/symmetry groups G}

Let us now explain in detail the relationship between SPT phases in
\Rf{CGL1172} and the topological lattice gauge theory discussed above.

In the construction in \Rf{CGL1172}, the theories are defined in $d$
dimensional space-time.  The wavefunctions with on-site symmetry group $G$ is
constructed making use of $d$ cocycles belonging to group cohomology group
$\cH^{d}(G, \R/\Z)$. For comparison with the topological lattice gauge
theories, we will for the moment restrict our attention to finite groups.
These cocycles have a geometric interpretation in terms of $d$ simplices.  To
write down the path integral, one considers a triangulation of the $d$
space-time manifold.  Physical degrees of freedom are attached to the vertices
of the triangulation. For an SPT phase associated with a symmetry group $G$, a
group element $g_{v_a} \in G$ is assigned to each vertex $v_a$. In addition,
the triangulation is endowed with a branching structure, exactly as the
topological lattice gauge theory described. The vertices are thus ordered on
each $d$ simplex $T^i_{d}$, giving orientations to each edge, and
subsequently an orientation for the $d$ simplex itself.  To each
$T^i_{d}$ it is attached a phase $\nu_{d}^{s}$, which is a function of
$d+1$ elements, taking values from each of the vertices of the $d$ simplex.
The orientation of the $T^i_{d}$ determines $s = +/\dag$, exactly as in
section \ref{sec:finitegauge}.  These $\nu$'s satisfy the following symmetry
relation: \be g\,.\nu_n(g_0,g_1,\cdots, g_{n}) = \nu_n(gg_0, gg_1,\cdots, gg_n)
\ee It has been shown \cite{CGL1172} that $\nu_n$ is indeed an $n$-cocycle, and
can be related to a more conventional form $\alpha_n$ by \be\label{mu2nu}
\alpha_{n}(g_1,\cdots, g_{n}) = \nu_{n}(g_1g_2\cdots g_{n},g_2\cdots
g_{n},\cdots, g_{n-1}g_{n}, g_{n},1).  \ee The path integral is then given by
\begin{align}
\label{SPTpath}
Z_{\textrm{SPT}} = |G|^{-N_v}\sum_{\{g_{v_a}\}} \prod_k
\left(\nu_{d}(\{g_{v_a}\}_{T_{d}^k})\right)^{\epsilon_k},
\end{align}
where the
product goes through all the $d$ simplices of the triangulation, and the sum
is over all possible sets of $\{g_{v_i}\}$ assigned to all the vertices. $N_v$
is the total number of vertices in the triangulation, and $|G|$ is the rank of
the group.

\begin{figure}[tb] \begin{center} \includegraphics[scale=0.5]{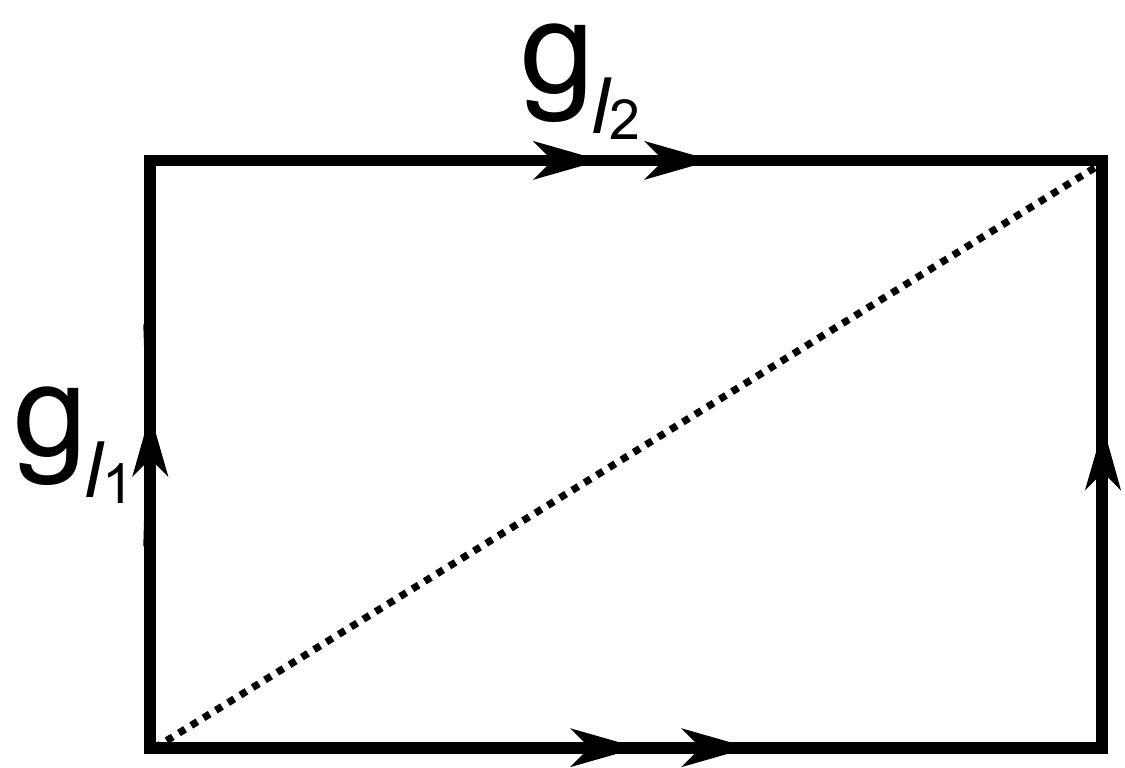}
\end{center}
\caption{A torus represented by a rectangle with opposite
sides identified. Each independent loop of the torus is assigned
a group element $g_{l_1}$ and $g_{l_2}$ respectively.} \label{torus} \end{figure}

This should be contrasted with the Chern-Simons path integral where physical
degrees of freedom $g_{ab}$ reside on each edge that connect nearest neighbor
vertices $v_a$ and $v_b$, and that each edge possesses an orientation, as
already explained in the previous section.  The relationship between the two
theories are straightforward:

\be\label{relation} g_{ab} = g_{v_a} g_{v_b}^{-1}.  \ee Note that this is a
$|G|^{n_0}$ to one map from the SPT phase to the lattice gauge theory, where
$n_0$ is the number of disconnected components.  Consider a single connected
component, and multiply each $g_{v_a}$ by a common group element $g$: i.e. take
$g_{v_a}\to g g_{v_a}$.  This gives rise to the same $g_{ab}$.  This
relation has important implication. Eqn. (\ref{relation}) provides a solution
to the flatness condition Eqn. (\ref{vanishflux}). It enforces that the flux
penetrating any triangle vanishes. However, one can compare with the gauge
transformation property of $g_{ab}$ that (\ref{relation}) implies that $g_{ab}$
is pure gauge. In a manifold where all loops are contractible, e.g. $S^n, n>1$,
indeed the only solution to the flatness condition is given by Eqn.
(\ref{relation}).  However, in general, when we have non-contractible loops,
there are extra solutions, corresponding to non-trivial \emph{Wilson lines}.
Since this is an important issue that underlies the fundamental difference
between the lattice gauge theory and the SPT phase, we illustrate the point by
giving a simple example and consider a two torus as in Fig. \ref{torus}.
We can represent it by a
rectangle, with opposite edges identified.  A simple triangulation would be to
introduce an extra edge on the diagonal of the rectangle. Now any one edge of
each of the triangles is a loop, since all the vertices of the rectangle are
identified. According to (\ref{relation}), we therefore have $g_{aa} = 1$ as
the only possible configuration.  However, from the point of view of the gauge
theory, it is also possible to assign non-trivial values to the sides of the
rectangle. Explicitly, considering one of the two triangles that make up the
rectangle, and label the states of the two external edges by $g_{l_1},
g_{l_2}$. The flatness condition requires only that the diagonal is restricted
to take values equal to $g_{l_1} g_{l_2}$, and that for consistency between the
two triangles sharing the same diagonal, $g_{l_1}$ and $g_{l_2}$ has to
commute.

Recall that in the lattice gauge theories, $d$-cocycles $\alpha_{d}$ arise
naturally in the fixed point wavefunction. These can be related to those
$d$-cocycles $\nu_{d}$ defined in \Rf{CGL1172} discussed above. The fact
that $\nu_{d}$ is a function of $d+1$ group elements, and that it is
invariant under the multiplication of all group elements by the same $g$ is
precisely the statement that $g_{ab}$ is insensitive to this transformation,
and $\alpha_{d}$ is invariant.

The $d$-cocycles $\alpha_{d}$ arising in the topological lattice gauge
theories are identified with $\nu_{d}$, except that the input values of the
function has to be translated according to Eqn. (\ref{relation}).  Making the
substitution into $\alpha_{d}$ reproduces the known relation between
$\nu_{d}$ and a $d$ cocycle as described above in Eqn. (\ref{mu2nu}).  We
recall that a $d$ cocycle is geometrically defined on the $T^{d}$ simplex,
which is topologically equivalent to a $d$ ball, whose loops are
contractible.  Therefore the lattice gauge theory and the SPT phase agree
there.

Finally we can compare the path-integral of the lattice gauge theory with Eqn.
(\ref{SPTpath}).  For the  SPT path integral, it is known that before taking
the overall sum over states, the action itself is equal to one in all closed
manifold.  Therefore the path integral is simply equal to the total number of
configurations, given by ${|G|^{N_v}}$, where $N_v$ is the total number of
vertices.  This can be interpreted roughly as the volume of the manifold. If a
state contains long-range entanglement, a non-trivial constant term independent
of volume should be expected, encoding information, for example about ground
state degeneracies.  Therefore the SPT phase is considered topologically
trivial.  Now, on a three manifold with $n_0$ disconnected components, but that
all the loops are contractible, the field configurations between the theories
are simply related by the $|G|^{n_0}$-to-1 map, and $Z_{\textrm{SPT}}$ is equal
to $ \mathcal{N} Z_{CS}$ up to volume independent choice of normalization  factor
$\mathcal{N}$.
\footnote{Our
choice of normalization for the lattice gauge theory in Eq. (\ref{pathI})
gives $\mathcal{N} = |G|$. }
The situation, however is more non-trivial if we begin to consider
manifolds with non-contractible loops.  In those cases, the lattice gauge
theories would then involve a sum also over Wilson lines. We will give some
simple examples of these non-trivial factors in the example sections.

While we have been discussing a duality relation between SPT phases and lattice
topological gauge theories (or Chern-Simons theories) in the case for finite
discrete groups $G$, we note that SPT phases can still be defined for
continuous $G$.  In that case, $\nu_{d}$ are not simple smooth functions of
the group variables, but instead are chosen to be \emph{Borel measurable
functions}\cite{CGL1172}.  They are classified by Borel group cohomology
$\cH_B^{d}(G, \R/\Z)$.  As discussed earlier,
while it is not immediately clear what the
corresponding lattice gauge theories should be except when the connection
is sufficiently close to vanishing, the Chern-Simons
theories constructed from differential characters are classified by
$H^{d+1}(BG, \Z) \cong \cH_B^{d}(G, \R/\Z)$. Therefore the correspondence
between Chern-Simons theories and SPT phases persists even for continuous $G$
in all space-time dimensions.

\section{Duality Relations with String Net models}

In Levin-Wen $2+1$
dimensional string-net models, each model is uniquely defined by a set of data:
$\{\mathcal{F}, F, N, \{d_g\}\}$, where $N$ is the number of string types in
addition to the trivial state, $\mathcal{F}$ denotes the fusion rules of the
string states when three meet at a vertex, and $F$ are the $6j$-symbols that
control crossing relations between different orders of fusing three states
$I,J,K$. (i.e.  $(I\otimes J) \otimes K$ verses $I\otimes (J \otimes K)$.)
Ultimately it also controls the form of the Hamiltonian. The quantum dimension
$d_I$ is assigned to each string type.

Each state of the system is defined on a 2 dimensional (closed) surface. String
states reside on edges of a trivalent lattice i.e. each vertex is connected to
three edges. The most studied example is the honeycomb lattice, although in
principle the lattice can be irregular.  The string state is labeled by a
representation of a group.  In the case of Abelian groups, representation of
the group is in 1-1 correspondence with group elements and thus
interchangeable. In the case of non-Abelian groups it is also possible to
perform the analogue of Fourier transformation to obtain a dual description in
terms of group elements\cite{BA0936}.  We will therefore for simplicity focus
our discussion on Abelian groups such that there is no real distinction between
group elements and irreducible representations up to a rescaling by phases of
the basis states, and we can use the two basis interchangeably. A
generalization to general groups would require following the procedure set out
in \Rf{BA0936} carefully.  In the following therefore, each edge of the
string-net lattice is associated with a group element $g_{e} \in G$ for some
Abelian group $G$.  An orientation is also assigned to each edge, but the
labeling is redundant-- the same state can be described by the element
$g_{e}^{-1}$ if the orientation is reversed.  At each vertex only three strings
meet.  The fusion rule then dictates that the incoming string states that meet
at a vertex product to identity.

To each configuration of the string-net, we associate to it a wavefunction
$\Phi$.  These wavefunctions satisfy a set of local rules, which are postulates
motivated by the physical requirement that they describe states that are at
fixed points of the renormalization group flow.  These local rules can be
summarized as follows:

\bse\label{eq:SNwaveFunc}
\begin{align}
\Phi\bpm\shadowGraph\hspace{-0.3pt}\sline\hspace{-1pt}\shadowGraph\epm
&=\Phi\bpm\shadowGraph\hspace{-0.3pt}\curveline\hspace{-1pt}\shadowGraph\epm
\label{eq:SNwaveFuncDeform}\\
\Phi\bpm\shadowGraph & \loopGraph{i}\epm
&=d_i\Phi\bpm\shadowGraph\epm\label{eq:SNwaveFuncLoop}\\
\Phi\left(\bmm\vspace{-3pt}\bubbleGraph{i}{j}{k}{l}\emm\right)
&=\delta_{ij}\Phi\left(\bmm\vspace{-3pt}\bubbleGraph{i}{i}{k}{l}\emm\right)
\label{eq:SNwaveFuncBubble}\\
\Phi\left(\bmm\vspace{-3pt}\scalefont{0.6}\Hgraph{j}{m}\emm\right)
&=\sum\limits_nF^{j_1j_2m}_{j_3j_4n}\Phi\left(\bmm\vspace{-3pt}\scalefont{0.6}\Xgraph{j}{n}\emm\right),
\label{eq:SNWaveFuncCross}
\end{align}
\ese

\begin{figure}[tb] \begin{center} \includegraphics[scale=0.5]{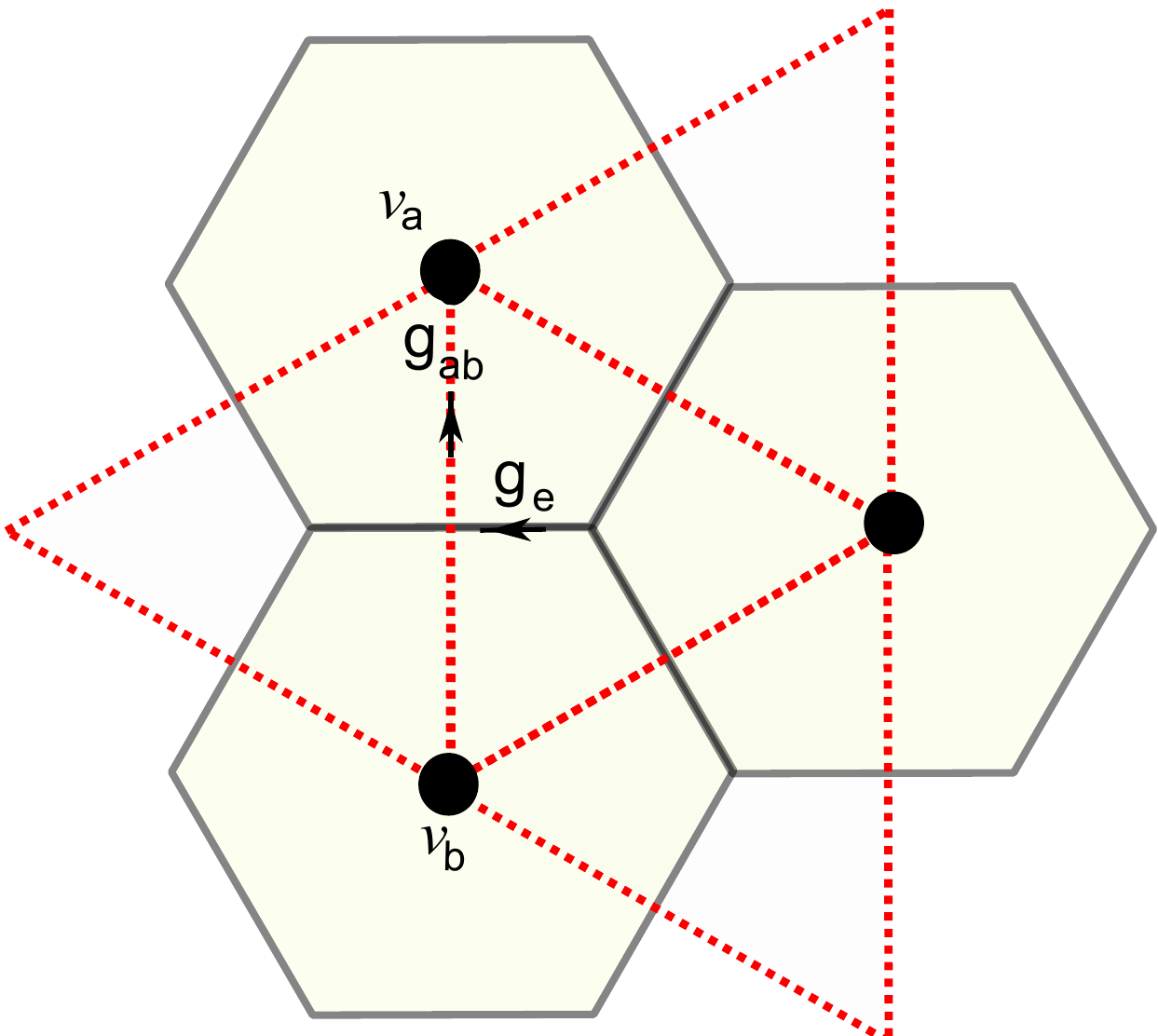}
\end{center}
\caption{String net on a honeycomb lattice and the dual triangular lattice
on which the lattice gauge theory is defined.} \label{duality} \end{figure}

The duality relation with the topological lattice gauge theory is closely
related to the duality relation already discussed in \cite{LG1220} between
the string-net model and the SPT phase. We note however that for the three
dimensional topological lattice gauge theory, the triangulation is taken over
the entire 3$d$ space-time manifold. The string-net state we have described
above, however, is understood as a state at a particular time in Hamiltonian
formulation. Therefore, it is not hard to guess that the wavefunction $\Phi$ is
related to the topological lattice gauge theory path integral with a boundary.
The relation between wavefunction and path-integral is well known in the
context of (conformal) quantum field theories. The wavefunction is, up to
normalization, understood as a path integral that integrates over all the paths
connecting a state from $t=-\infty$ to some state at finite $t_0$. A state at a
given time-slice really means boundary conditions on the fields.  In the
lattice gauge theory path-integral, the gauge fields $g_{ij}$ also satisfy
fixed boundary conditions respected by the path integral.  A given
configuration of these boundary degrees of freedom denoted by
$\Gamma_\textrm{dual}$ can be mapped to a unique string-net state $\Gamma$. To
make the map precise, the 2-dimensional fixed time slice inherits a
triangulation from the triangulation of the entire 3-manifold, and a group
element is again attached to each edge of the triangles. This triangular
lattice is a dual lattice of the honeycomb lattice.  i.e. The triangular
lattice is given by the set of vertices that reside at the center of each
hexagon of a given honeycomb lattice, as shown in Fig. \ref{duality}.

 Each edge on the
honeycomb lattice therefore cuts through precisely one edge on the triangular
lattice that connects two nearest neighbor vertices. This suggests that one
should identify the
string state $g_e$ on an edge $e$ and the gauge state on the edge of the
triangular lattice $g_{ab}$ intersecting $e$. The orientation of the
string state $g_e$ in the duality can be chosen by convention, as depicted
in Fig. \ref{duality}.
Ultimately the duality is such that a face is mapped to a vertex and vice versa,
whereas an edge is mapped back to an edge. The honeycomb lattice
serves only as an example to illustrate the relation, but the relation
survives independently of the precise choice of the lattice, as
long as it is trivalent. This is necessary since the local rules
described in Eq. (\ref{eq:SNwaveFunc}-\ref{eq:SNWaveFuncCross}) do not generally
preserve the lattice structure.

The explicit relation between the string-net wavefunction $\Phi(M_2, \Gamma)$
on a closed 2d surface $M_2$ for a given string-net configuration $\Gamma$,
and
the path integral $Z_{CS}(B, \Gamma_\textrm{dual})$ of the topological lattice
gauge theory on a 3d manifold $B_3$ where $\partial B_3 = M_2$ with
corresponding boundary gauge configurations $\Gamma_\textrm{dual}$,
is then given by
\begin{align}
\label{phiDW}
&\ \ \ \Phi(M_2)[\Gamma] =\frac{
\mathcal{N}[\Gamma]}{|G|^{N_{v_\textrm{int}}}} Z_{CS}(B_3)
\\
&=
\frac{ \mathcal{N}[\Gamma]}{|G|^{N_{v_\textrm{int}}}} \sum_{\{g_{ab}\}_{a,b\in
\{v_\textrm{int}\}}} \prod_{i}\alpha^{s_i}(\{g_{c d}\}, \{v_{c,d} \in T_i\}),
\nonumber
\end{align}
where $\{v_\textrm{int}\}$ denotes the set of
vertices lying in the interior of $B_3$, and $N_{v_\textrm{int}}$ is the total
number of these internal vertices, and $ \mathcal{N}[\Gamma]$ is some
normalization that depends on the specific string-net configuration, or surface
gauge configuration, denoted by $\Gamma$.  This corresponds to the freedom in
the string-net model to choose a basis set of wavefunctions for each
configuration. The wavefunctions are however chosen to be related to each other
by specific rules \Rf{LG1220}, which has specific meanings in the lattice
gauge theory as we will explain below.  Using the duality relations, we note
immediately that $\alpha( \{g_{ab}\})$, the action defined on a single
tetrahedron, is proportional to the string-net wavefunction also on a
tetrahedron.  Given the cocycle condition satisfied by $\alpha(\{g_{ab}\})$,
the wavefunction is clearly independent of the precise triangulation in the
interior of the manifold $B_3$.

\begin{figure}[tb] \begin{center} \includegraphics[scale=0.9]{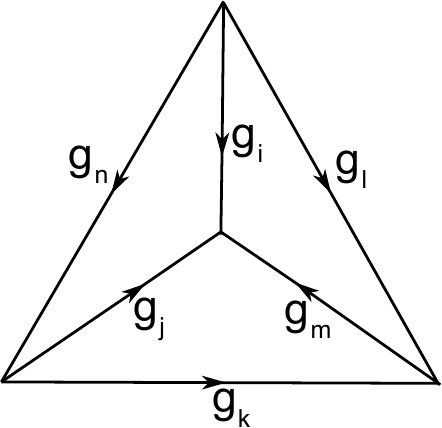}
\end{center}
\caption{The string net dual of a tetrahedron which is still a tetrahedron.
} \label{sntetra} \end{figure}

In the normalization chosen in \Rf{LG1220}, the wavefunction of a string-net
on the tetrahedron as depicted in Fig. \ref{sntetra} is given by \be \Phi(T)=
v_{g_i} v_{g_j}v_{g_k} v_{g_l} F^{g_ig_jg_m}_{g_kg_lg_n}, \ee where
$F^{g_ig_jg_m}_{g_kg_lg_n}$ is a component of the $6j$-symbol, and $v_{g_i}^2
\equiv d_{g_i} =\pm 1$ for Abelian groups.  The six elements involved are not
independent due to fusion rules. In fact \be g_m=g_k  g_l\,,\qquad g_n=
g_kg_j\,,\qquad g_ig_jg_kg_l=1.  \ee The $6j$ symbols satisfy the pentagon
relations, and that the pentagon relations are invariant under a rescaling of
$F$'s, corresponding to a rescaling of each of the vertices. i.e.  \be
F^{g_ig_jg_m}_{g_kg_lg_n} \to \frac{f(g_i,g_l,g_n) f(g_j,g_k,g_n^{-1})}
{f(g_i,g_j,g_m) f(g_k,g_l,g_m^{-1})} F^{g_if_jg_m}_{g_kg_lg_n}.  \ee where the
phase factor $f(g_i,g_l,g_n)$ is symmetric under cyclic rotation of the three
elements, and recall that of which only two are independent since the fusion
constraint requires that $g_i g_l g_n=1$.  We therefore identify the set of
$6j$ symbol as a 3-cocycle in $H^3(G,U(1))$ in group cohomology, and that the
above rescaling is a rescaling of a 3-cocycle by a coboundary $\delta\beta$ for
some 2-cochain $\beta$. i.e.  \be\beta(g_{i},g_l)=f(g_i,g_l,g_n).  \ee
Carefully comparing the transformation property of
$\alpha(g_{ab},g_{bc},g_{cd}), \, a<b<c<d $ under rescaling by a co-boundary we
therefore have \be v_{g_i} v_{g_j}v_{g_k} v_{g_l} F^{g_if_jg_m}_{g_kg_lg_n} =
\mathcal{N}[\{\Gamma_\textrm{tetrahedron}\}]\alpha(g_i,g_j,g_k), \ee and
therefore \be \label{faces} \mathcal{N}[\Gamma_\textrm{tetrahedron}]= v_{g_i}
v_{g_j}v_{g_k} v_{g_l}.  \ee

There is however one important distinction between the string-net model and the
topological lattice gauge theory. That is, the string-net wavefunction defined
on a tetrahedron is chosen such that it respects the tetrahedron symmetry, and
that the branching structure introduced in the lattice gauge theory is absent.
This imposes very severe constraint on the solution of the $6j$-symbols, and as
a result, the solutions considered in such highly symmetric string-net models
do not span the full $H^3(G,U(1))$, where this is also observed
for example for the explicit case of $G=\mathbb{Z}_3$ in \Rf{HW}.
Generalization via introducing a branching
structure is clearly possible, particularly given the duality relation with the
lattice gauge theory. Numbered vertices on the surface of the lattice gauge
theory is dual to plaquettes of the string-net lattice.  The branching
structure therefore admits a direct translation.\footnote{We thank Y. Wu, Y. Hu
and Y. Wan for communicating to us this fact to be published in their
forth-coming work. }

As mentioned above, the wavefunction $\Phi$ of a string-net configuration can
be related to some other configurations via a set of local rules. These local
rules have simple implementation from the the perspective of the path integral
of the topological lattice theory.
\subsection{Crossing relation}
\begin{figure}[tb] \begin{center} \includegraphics[scale=0.7]{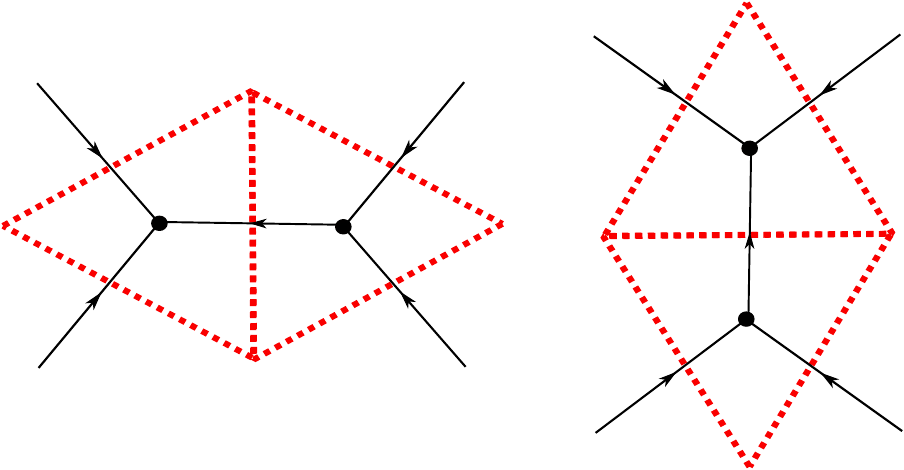}
\end{center}
\caption{Crossing relations in the dual lattice gauge theory (in dotted
red lines). } \label{crossdual} \end{figure}

Eq. (\ref{eq:SNWaveFuncCross})
is referred to as the crossing relation which
has a dual in the lattice gauge theory as
relating two sets of triangles, depicted as in Fig. \ref{crossdual}.
The crossing transformation on the wavefunction
corresponds to placing an extra tetrahedron $T$ right on top of the specific
triangles. i.e. two of the triangular surfaces of $T$ is matched/glued to the
two surface triangles involved.  Since there isn't any extra internal vertices
involved, this addition lead only to an extra factor $\alpha(\{g_{ab}\})$ in
the path integral, exactly as expected of the transformation property of the
string-net wavefunction $\Phi$.

\subsection{Removal of isolated loops} 
\begin{figure}[tb] \begin{center} \includegraphics[scale=0.5]{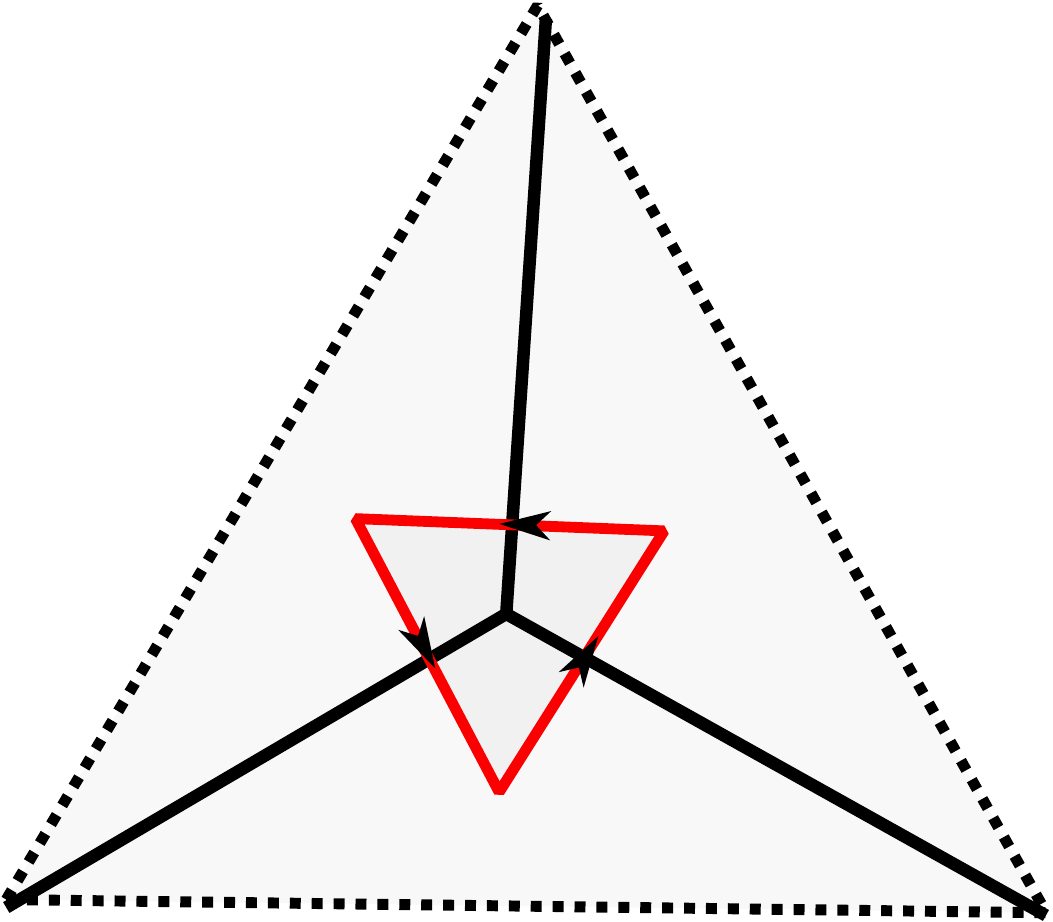}
\end{center}
\caption{A loop in the string-net model, and its dual in the lattice
gauge theory.
.} \label{loop} \end{figure}

In the string-net model, an isolated
loop of string type $i$ can be removed as the wavefunction acquires an extra
factor of $d_i$, the quantum dimension of the string type $i$. Consider a small
loop in the string-net model. It corresponds to three triangles meeting such
that together they form a larger triangle, as shown in Fig. \ref{loop}.

The gauge configuration is
such that the internal lines take values $i$, and the external edges of the
large triangle are in the trivial state. Topologically, that is precisely a
tetrahedron fitted at the surface. Since all the edges belong to the boundary,
there is no summation required over the states on the tetrahedron, and this is
a simple factor.  Therefore, we could replace the tetrahedron by its numerical
value. At the same time, an extra phase factor factor is attached accounting
for the removal of the three edges. This factor is chosen to be $d_i$, which we
recall is $\pm1$ for Abelian groups.  We note that any loop can be reduced via
multiple crossings to the basic loop involving the three triangles described
above.

We note that these rules uniquely determine the ground state wavefunction.
However, the wavefunction itself is not a topological object. If we were to
compute ground state degeneracy of the string-net on a closed 2d manifold
$\Sigma_g$ of genus $g$, it is given by the following: \be
\textrm{degeneracy}(\Sigma_g) = \sum_i \Phi[V_i]\Phi^{*}[V^i], \ee where $V_i$
denotes the basis for ground states. We recognize the above, given the
relationship Eq. (\ref{phiDW}), as gluing two path integrals, each of which
defined on 3-manifolds $\Sigma_g \times I$, for some finite interval $I$, along
the common boundary $\Sigma_g$. Since the normalization factor
$\mathcal{N}[\Gamma]$ is a phase, it is canceled out in the computation of
degeneracy, and we are left with a path integral of the topological lattice
theory over a closed 3-manifold $\Sigma_g \times S^1$, which is a topological
invariant.

\setcounter{subsubsection}{0}
\section{Some simple examples} In this
section, we would like to give some simple examples of these lattice gauge
theories. We would particularly be interested in finite gauge groups $G$  in
one and two spatial dimensions.

\subsection{$d$=3} To begin with we
will take the above construction and study the explicit form of the action of
the lattice topological gauge theory for a finite group $G$ at $d=3$, the case
which has already been discussed in detail in \Rf{DW9093}.

Recall that a triangulation of a three manifold $M_3$ is given by \be M_3 = \sum_i
\epsilon_i T_i, \ee where $T_i$ are 3 simplices, or in other words tetrahedra.
There are 4 vertices and 6 edges on a tetrahedron.  As already discussed in the
previous section, each edge connecting vertices $v_i$ and $v_j$
is assigned a group element $g_{ij}$.
To make the assignment unambiguous, one needs
to give an orientation to each edge, and thus a branching structure to the
tetrahedron. This is achieved by numbering the vertices, from 0 to 3.  Of the 6
group elements assigned to the 6 edges, only three of them are independent.  To
see that, we recall that each 2-simplex, or triangle, lead to one constraint
between the edges. However, these constraints are not independent. For a
tetrahedron, the fact that the triangles together form a closed surface (ie the
boundary of the tetrahedron) signals that there is one redundant constraint.
Therefore the number of independent degrees of freedom is given by
\begin{eqnarray} \textrm{dof} &=&  \textrm{\# edges} - (\textrm{\# triangles} -
\textrm{\# connected piece}) \nonumber \\ &=& C^4_2 - C^4_3 + C^4_4 = 3,
\end{eqnarray} $C^{d}_i$ are binomial coefficients.

Let us note that this result can be generalized easily to general dimensions.
\be \textrm{dof}_{d-\textrm{simplex}} = \sum_{i=2}^{d} (-1)^i C^{d}_i = d.
\ee

\subsubsection{Path-integrals for $\mathbb{Z}_N$}

\begin{figure}[tb] \begin{center} \includegraphics[scale=0.3]{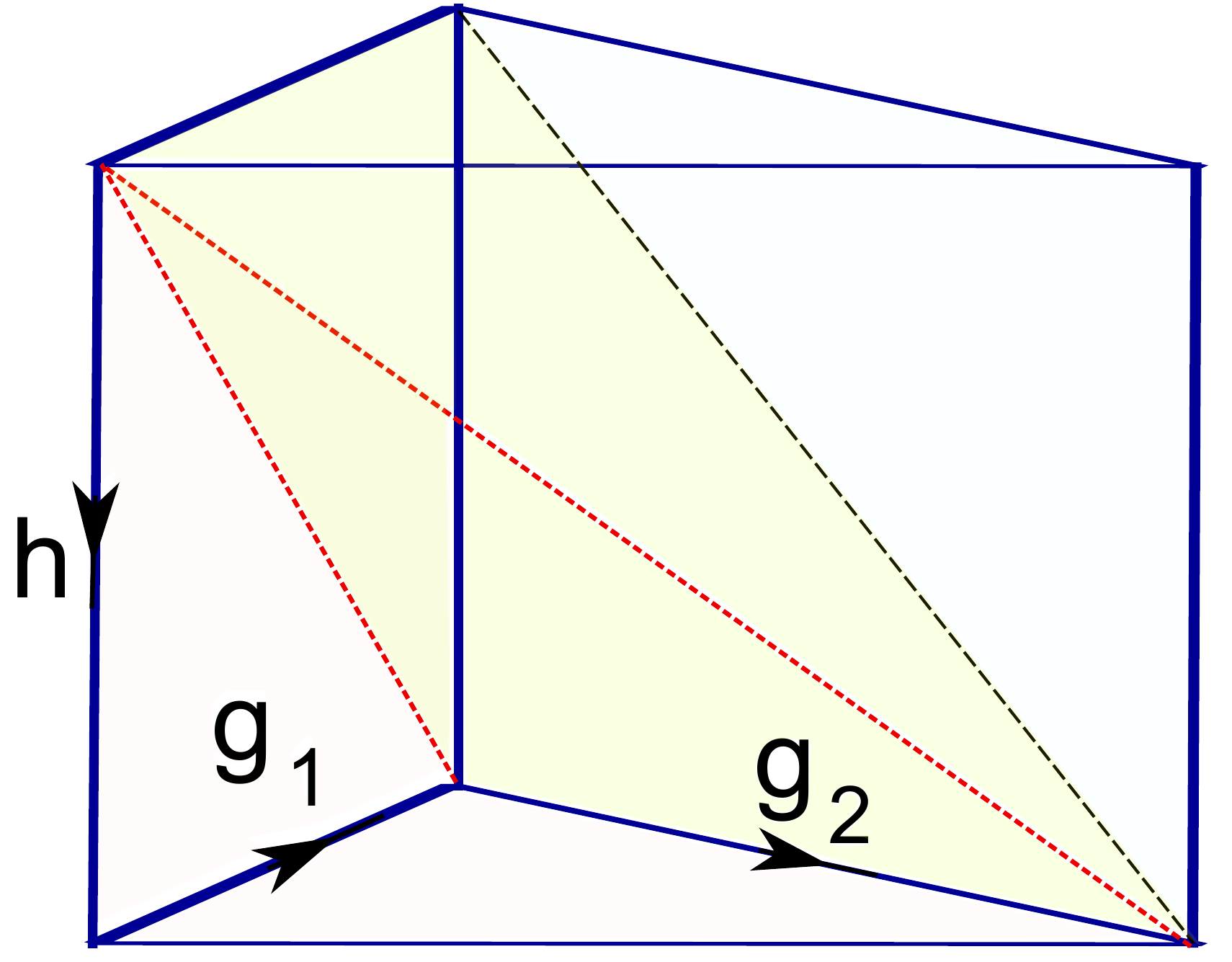}
\end{center}
\caption{The \emph{cellurarization} into three tetrahedrons of $Y \times S^1$.}
\label{Yblock} \end{figure}

To compute the path-integral of the lattice gauge theory on a general three
manifold, it is useful to first consider the special case where  the manifold
concerned is given by $Y \times S^1$, where $Y$ is a 2-sphere with
three holes. The triangulation is represented in Fig. \ref{Yblock},
where the sphere is represented by the triangle whose three vertices are
identified, and the holes by the three edges, and $S^1$ corresponds
to the vertical edge perpendicular to the triangle.

Strictly
speaking, this is not a triangulation but a \emph{cellularization} since
each edge connects to the same vertex. The naive branching structure obtained
by ordering the vertices become ill-defined, since there is only one
vertex. A rigorous triangulation
would require adding extra vertices to our present construction. However that
does not affect the final result of the path-integral given
its nature as a topological invariant. For simplicity as in
\Rf{DW9093}, we keep
to this simple \emph{cellularization}, but
replace the branching structure by explicitly specifying the
orientation of each edge.
A group element is assigned to each of the cycles, subjected to the flatness
condition on the triangle. The
consistency condition also immediately follows:
\be
[h,g_i]=1.
\ee
Again, the above condition is a result of
considering the group element assignment to diagonals on any of the
vertical rectangles.
Since the manifold is open, no summation is required over
the group elements, and the path-integral with the orientation
assignment as in the figure is given by
\be \label{ZY}
Z_{Y\times S_1} = \frac{1}{|G|}\frac{\alpha(h,g_1,g_2)\alpha(g_1,g_2,h)}{\alpha(g_1,h,g_2)}
\equiv  \frac{1}{|G|} c_h(g_1,g_2).
\ee
It is denoted $c_h(g_1,g_2)$ because it can be readily shown that it is
a 2-cocycle of the group $N_h$, where $N_h\subset G$ denotes the subgroup
whose elements commute with $h$.
From the relation between $c_h(g_1,g_2)$ and the 3-cocycles
$\alpha$, we should rewrite $|G| Z_{Y\times S_1}=c_h^{\epsilon}(g_1,g_2) $.
If the orientation of the triangle
aligns with that of the vertical edge we obtain $\epsilon= +1$, and
$\epsilon = \dag$ otherwise.

Path-integrals of more general three manifolds can
be obtained by gluing together $Z_{Y\times S_1}$ via
the gluing conditions Eq. (\ref{gluing}).

\begin{figure}[tb] \begin{center} \includegraphics[scale=0.20]{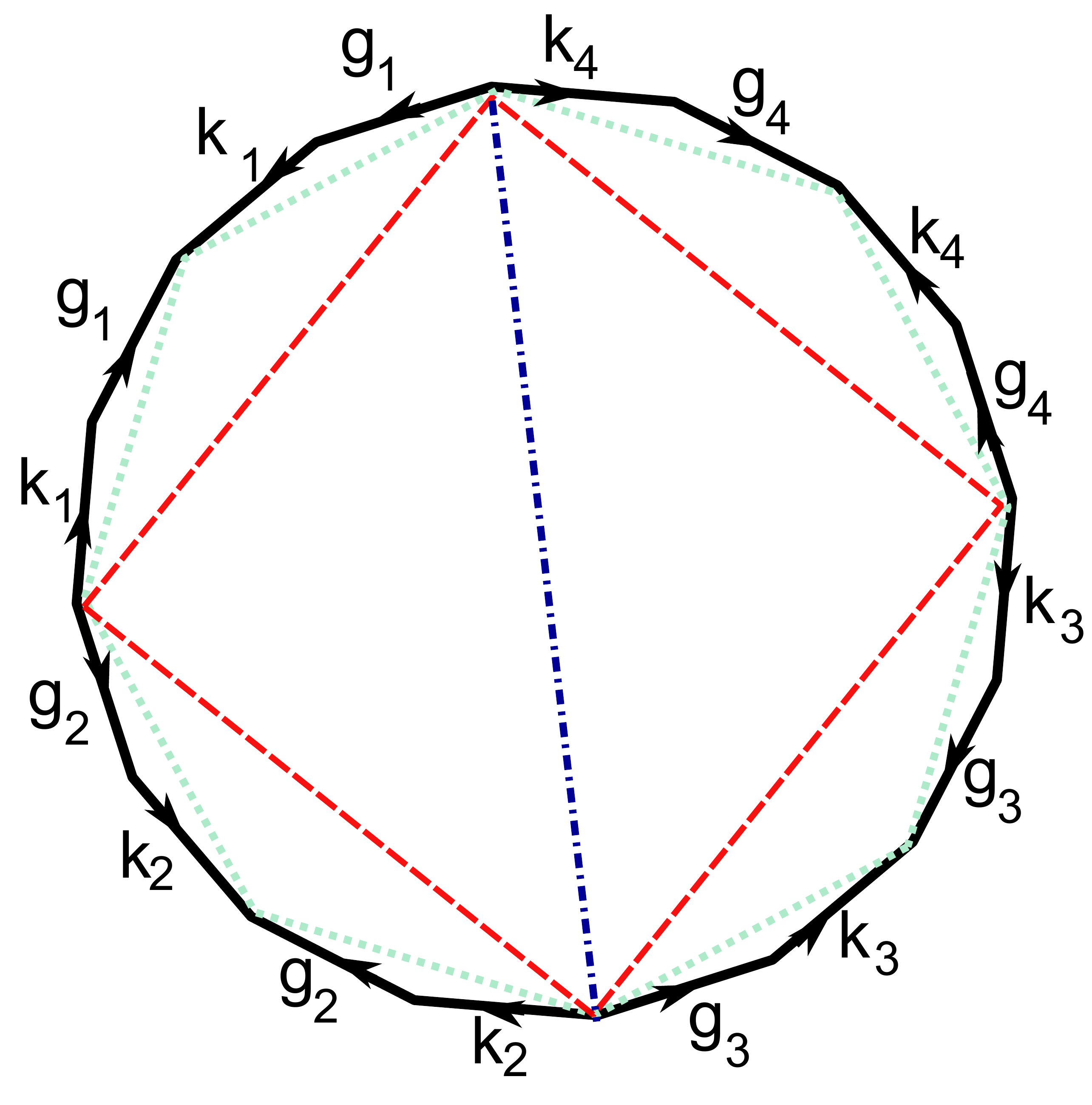}
\end{center}
\caption{A depiction of the genus four surface $\Sigma_4$. The extra circle
$S^1$ described in the text
is suppressed. Edges assigned
the same group element is also identified. The dashed lines correspond
to our canonical choice of dividing the three dimensional block into
$Y \times S^1$, which is then used for the computation of the partition function.
} \label{highg} \end{figure}

One important class of three manifolds are of the form $\Sigma_g \times S^1$,
where $\Sigma_g$ is a genus $g$ two dimensional closed surface. The path-integral
evaluated on these manifolds can be interpreted as ground state degeneracy of the
system residing in the spatial slice $\Sigma_g$.
The cellularization of a $\Sigma_g \times S^1$ is demonstrated
in Fig. \ref{highg}.

It is readily built up from a collection of $Y\times S^1$. A group
element is assigned to each non-trivial loop, and we therefore have the collection
$\{g_i,k_i\},\,\, 1\leq i\leq g$. For consistency
we again require that
\be \label{constr}
\prod^g_i[g_i,k_i]= 1\,,\qquad [h,g_i]=[h,k_i]= 1,
\ee
where $1\leq i \leq g$.
The path-integral is then given by a sum over all $h,g_i,k_i$
subjected to the above constraints,
\begin{eqnarray}
&&Z_{\Sigma_g\times S^1} = \frac{1}{|G|} \sum_{g_i,k_i,h}
 \prod_{i_1}^g \frac{c_h(g_{i_1},k_{i_1})}{c_h(k_{i_1},g_{i_1})} \nonumber \\
&&\prod_{i_2}^g c_h(g_{i_2}k_{i_2},(k_{i_2}g_{i_2})^{-1})
 \prod_{i_3}^{g-3} c_h(\prod_j^{i_3} x_j, x_{i_3+1} )
\label{Zsigmag}
\end{eqnarray}
where $x_i = [g_i,k_i]$. To understand the above form, we note
that the first product corresponds to the outermost set
of triangles bounded by the boundary of the polygon and the light
blue lines in Fig. \ref{highg}; the second product
corresponds to the set of triangles bounded between the light
blue lines and the red lines. Cutting along
the light blue and red lines give several blocks
of $Y\times S^1$, which contribute to the first two products.
Then we are always left
with a $g$-gon for a genus $g$ surface. The third set of products
then correspond to triangulating the remaining $g$-gon by
connecting a chosen vertex with all other vertices
except its own nearest neighbor. There are thus $g-3$ such vertices.
In Fig. \ref{highg} there is exactly $4-3=1$ such cut corresponding
to the deep blue line.

Consider specifically $\mathbb{Z}_N$ groups.
The 3-cocycles of $\mathbb{Z}_N$ is given by \cite{MS8977}
\be\label{ZN3cocycles}
\alpha_k(g_1,g_2,g_3) = \exp(\frac{2\pi i k \bar{g}_1}{N^2}(\bar{g}_2
+ \bar{g}_3 - \overline{(g_2+g_3)})),
\ee
for some appropriate $k \in \mathbb{Z}$, and $g_i \in \mathbb{Z}_N$,
and $\bar{x} = x \,\,\,\textrm{mod} N$ for $x\in\mathbb{Z}$. There are
altogether $N$ distinct choices of $k$ that give rise to representatives
of the $N$ different group cohomology classes in $H^3(\mathbb{Z}_N,U(1))$.
However, substituting the above expression into Eq. (\ref{Zsigmag})
the summands become identically 1, independently of $k$!
Taking also into account that the group is Abelian and thus the
constraints in Eq. (\ref{constr}) are trivially satisfied,
the sum simply separately counts the possible Wilson
loop around each of the $2g$ cycles in $\Sigma_g$. Therefore, we have
\be
Z_{\Sigma_g\times S^1, G= \mathbb{Z}_N} = N^{2g}.
\ee
This is a special case of the result obtained in \Rf{DW9093} for more general
finite groups $G$.
We note that at the end this appears to have no dependence
on the 3-cocycle $\alpha$ we have chosen. The reason is
that these manifolds $\Sigma_g\times S^1$ can
all be computed by cutting them into blocks of $Y\times S^1$. As
we have seen in Eq. (\ref{ZY}) the 3-cocycles always
come in combinations to form $c_h(g_1,g_2)$. These are classified
by $H^2(N_h,U(1))$. For
Abelian groups these are trivial, and for more general
groups, they are in general different from $H^3(\mathbb{Z}_N,U(1))$.
Since our path-integral on $\Sigma_g\times S^1$ is equal to the ground state
degeneracy, it is perhaps not surprising
that the ground state degeneracy alone does not distinguish all
topological phases.

\subsubsection{The $S$ and $T$ matrix}

There are two other important quantities required to obtain
the partition function on general 3-Manifolds. These are the so called
$S$ and $T$ matrices which describe how the partition function
on a 3-manifold $M$ whose boundary is a torus transforms under
modular transformation of the torus.
So far we have been working with a basis for the Hilbert space
given by different configurations of
group elements assigned on the edges residing on
the boundary of $M$. To make closer connections with Chern-Simons
theory, it is useful to make a change of basis and label the states
in terms of representations. Note that this change
of basis is exactly the same transformation that
connects the lattice gauge theory and the string-net models
discussed in the previous section.  In such a basis, through the relationship
of the lattice gauge theory with 2-dimensional orbifold theories,
the $S$ matrix is  known to be\cite{DVV8985}
\be\label{Smatrix}
S^{AB}_{\alpha\beta} = \frac{1}{|G|}\sum_{h\in C_B, g\in C_A, [h,g]=1} \rho^g_{\alpha}(h^{-1})
\rho^h_{\beta}(g^{-1})\sigma(g\vert h),
\ee
where $C^{A,B}$ denote conjugacy classes of the group $G$ and
$\rho^g_\alpha = \textrm{tr} R^g_\alpha$, where $R^g_\alpha$ is a representation of
the stabilizer group $N_{g\in C^A}$ containing elements that commute with $g$. The
subscript $\alpha$ enumerates these representations for each $N_{g\in C^A}$.
Since $N_{g}$ is isomorphic to each other for all $g$ belonging to
the same conjugacy class, $N_g$ is equivalently denoted as $N_A$.
Correspondingly, the $T$ matrix is given by
\be\label{Tmatrix}
T^{AB}_{\alpha\beta}= \delta_{\alpha\beta}\delta^{AB}
\rho^g_\alpha(g)\rho^g_\alpha(1)^{-1} \sigma(g\vert g)^{-1/2}.
\ee
For Abelian groups, $H^2(G, U(1))$ is trivial.
The phase $\sigma(g\vert h)$ treated as a 1-cochain in $N_h$ is
then related to $c_h(g_1,g_2)$ by
\be \label{sigmagh}
c_h(g_1,g_2) = \frac{\sigma(g\vert h_1)\sigma(g\vert h_2)}{\sigma(g\vert h_1h_2)}.
\ee

Consider $\mathbb{Z}_N$. In this case there are $N$ distinct
classes in $H^3(\mathbb{Z}_2,U(1))$, with representative
as already given in Eq. (\ref{ZN3cocycles}).
Using Eq. (\ref{ZY}, \ref{sigmagh}), we can read off $\sigma(g\vert h)$.
They are given by
\be
\sigma_{(k)}(g\vert h) = \exp(\frac{2\,i \pi k g h}{N^2}).
\ee
Also all stabilizer subgroups $N_A= \mathbb{Z}_N$.
For concreteness, let us consider $N=2$. In which case $k=0$ or $k=2$.
As a result $\rho^g_{\alpha}(h) = (-1)^{\alpha h} $, where both $\alpha$
and $h$ can take values $\pm1$.
Substituting into Eq.(\ref{Smatrix}, \ref{Tmatrix}), and writing them
as $4\times 4$ matrices, we have the following two sets of $S,T$ matrices:
\begin{eqnarray}
T (k=0)&&= \textrm{diag}(1,1,1,-1), \\
S(k=0)&&= \frac{1}{2}\left(\begin{array}{cccc}
1&1&1&1 \\
1&1&-1&-1 \\
1&-1&1&-1 \\
1&-1&-1&1
\end{array}\right),
\end{eqnarray}
and
\begin{eqnarray}
T(k=2)&&= \textrm{diag}(1,i,-i,1), \\
S(k=2)&&= \frac{1}{2}\left(\begin{array}{cccc}
1&1&1&1 \\
1&-1&1&-1 \\
1&1&-1&-1 \\
1&-1&-1&1
\end{array}\right).
\end{eqnarray}
These matrices are precisely those $S$ and $T$ matrices obtained in \Rf{LW0510}
for the two distinct $\mathbb{Z}_2$ string-net models. As discussed there,
these $S$ and $T$ matrices also arise from $U(1)\times U(1)$ Chern-Simons theories.
The $k=0$ $\mathbb{Z}_2$ lattice gauge theory is equivalent to the $U(1)\times U(1)$
Chern-Simons theory with $K$-matrix
\be
K_0= \left(\begin{array}{cc}
0&2 \\
2&0\end{array}\right),
\ee
whereas the $k=2$ theory corresponds to
\be
K_2= \left(\begin{array}{cc}
2&0 \\
0&-2\end{array}\right).
\ee
Ironically, the correspondence underlies the fact that a gauge group in a gauge
theory is not really a physical object, but rather a tool in constructing
redundancy in a theory.

\subsection{$d$=2}

Having discussed the case of $d=3$, let us
also look at some simple examples in $d=2$.
Our discussion in this paper so far has refrained from
including $d=2$. The reason is that in 2 dimensions it is well known that
long range order is necessarily destroyed by strong quantum fluctuations.
In 2$d$ gapped systems without symmetries, which is the case of a weakly coupled gauge
theory necessarily belong to the same phase\cite{CGL1172}.
We believe the phases studied
here is ultimately unstable.
More specifically, the flatness condition crucial
to these phases probably does not survive quantum fluctuations.
We study examples in one spatial dimensions to illustrate some features of
these lattice gauge theories and their corresponding partition function, in the
hope that some of the characteristics of the action can be extrapolated
in higher dimensions.

In
this case a triangulation of a two manifold $M_2= \sum_i \epsilon_i T_i$ divides
the manifold into triangles $T_i$.
Each triangle would then be associated with a 2-cocycle $\alpha_2 \in H^2(G,U(1))$.
For a general closed orientable genus $g$ manifold $\Sigma_g$, we
can adopt basically a very similar cellularization as in the
$d=3$ case where $M_3 = \Sigma_g \times S^1$,
and the final
result of the path-integral is simply given by the same expression (\ref{sigmagh}),
except that each $c_h(g_i,k_i)$ is replaced by $\alpha_2(g_i,k_i)$, and that there
is one less summation over Wilson loops along the extra $S^1$ there.
For Abelian groups, $H^2(G,U(1))=0$, and we can simply take $\alpha_2 (g,k)\equiv 1$.
Therefore, the result is simply given by
\be \label{Z2dA}
Z_{\Sigma_g} = |G|^{2g-1}.
\ee

We would like to pause here and comment on the case where $G$ contains
time-reversal $\mathbb{Z}^T_2$.  In this case, all group elements are
divided into two groups, such that each element $g\in G$
is assigned a value $t(g)=\pm1$. The $n$-cocycle condition is modified to
\begin{eqnarray}
&&(d_n\alpha_n)(g_1,\cdots,g_{n+1}) =\nonumber \\
&&\alpha_n^{t(g_1)}
(g_2,\cdots,g_{n+1})\alpha_n^{(-1)^{n+1}}(g_1,\cdots,g_n) \nonumber \\
&&\prod_i^n \alpha_n^{(-1)^i}(g_1,\cdots,g_{i-1},g_if_{i+1},g_{i+2},\cdots g_{n+1}).
\end{eqnarray}

SPT phases with time-reversal symmetry has been
discussed already in \Rf{CGW1038}. Here, following completely
analogous construction of the lattice gauge theory, we can define a path-integral
for a lattice gauge theory which gauges time reversal symmetry!
The path-integral is given by
\begin{eqnarray}
&&Z_M({G \supset \mathbb{Z}^T_2}) =  \nonumber \\
&&|G|^{-N_v}\sum_{\{g_{ab}\}}\prod_{i}
\alpha^{s_i * t(g_b)}(\{g_{cd}\},\{v_{c,d}\in T_i\}).
\label{trevZ}
\end{eqnarray}
Here $s_i$ depends on the orientation of the $d$-simplex $T_i$ as explained
in section \ref{ltgauge}. The new ingredient is $t(g_{p_i})$. Recall that we have
assigned a branching structure to the triangulation, giving an
order to the vertices, both locally on each simplex, and globally.
This defines a special base point $P_0$ in the space-time manifold, i.e. the vertex
with the smallest index, and also a base point $p_i$
in each simplex. Consider a path connecting $P_0$ and $p_i$, passing through vertices
$v_1,v_2,\cdots v_n$, then we define
$g_{p_i}= g_{P_0 v_1} g_{v_1 v_2}\cdots g_{v_n v_{p_i}}$.
The value $t(g_{p_i})$ is then $+1$ or $\dag$ depending on $g_{p_i}$.
Note that $t(g_1 g_2)= t(g_1)t(g_2)$, therefore the assignment of $t(g_{p_i})$
to each simplex is independent of the path chosen to connect the base points.
Homotopic paths would automatically give the same
$t(g_{p_i})$.

We would like to compute the path-integral for the simplest
case where $G= \mathbb{Z}_2^T$ for $d=2$. In this case,
$H^2(\mathbb{Z}_2^T, U(1))= \mathbb{Z}_2$. The explicit form of
representative cocycle from each class of $H^2(\mathbb{Z}_2^T, U(1))= \mathbb{Z}_2$
has been computed in \Rf{CGW1038}. We represent $\mathbb{Z}_2^T = \{0,1\}$,
and we take the only non-trivial consistent
choice of $t(0)=1$, and $t(1)=\dag$. Group multiplication
is then taken as addition of group elements modulo 2.
By making use of
rescalings using coboundaries, it is demonstrated\cite{CGW1038} that it is convenient to
exhaust the redundancy by rescaling all components to 1,
leaving behind only $\alpha(1,1)$.
A representative of each of the classes is then given by $\alpha(1,1)=\pm1$.
Therefore, we can rewrite the two classes as
\be
\alpha_{\pm}(g_1,g_2)= \exp(\frac{i\pi (1\mp 1)}{2} g_1  g_2).
\ee

The fact that these 2-cocycles can be made completely real actually
tells us that the extra group action $t(g_{p_i})$ attached
to each $\alpha$ on each simplex does not lead to anything new in this
particular case.
Substituting these $\alpha$ into the path-integral Eq.(\ref{trevZ})
should yield the same answer as $\mathbb{Z}_2$ for arbitrary
closed orientable surface of genus $g$. Indeed, by explicit
substitution, and the fact that $\sum_i^g (g_i+k_i)=0$,
we recover the result in Eq. (\ref{Z2dA}) with $|G|=2$.

\section{Summary}

In this paper, we consider weakly coupled lattice gauge theories where the charged
particles have a large mass gap and the field strength fluctuations are weak at
the lattice scale.  A weakly coupled lattice  gauge theory is described by the following
imaginary-time path integral:
\be
\label{Zgauge}
Z =
\sum_{\ga} e^{\imth S_\text{top}[\{G_{ij}\}]-S_\text{dyn}[\{G_{ij}\}]}.
\ee
where $S_\text{top}[\{G_{ij}\}]$ is a topological term which is invariant under
the coarse graining of lattice and is independent of ``space-time metrics''.
The dynamical term $S_\text{dyn}[\{G_{ij}\}]$ imposes the conditions of the
large mass gap for the charged particles and the weak fluctuations for the
field strength.  We find that the \emph{quantized topological terms} can be
constructed systematically from the elements of topological cohomology classes
$H^{d+1}(BG,\Z)$ for the classifying space of the gauge group $G$ in $d$
space-time dimensions.  This result is valid for any compact gauge groups
(continuous or discrete) and in any dimensions.  This generalizes the
Chern-Simons topological terms and other form of topological terms previously
known in gauge theory.

Our motivation to study quantized topological terms is to gain some general
understanding of quantum phases of gauge theory.  Since quantized topological
terms cannot be modified under the renormalization flow, it is possible that
adding a quantized topological term will cause the system to go to another
phase.  In $3$ space-time dimensions, the quantized topological terms
correspond to generalized Chern-Simons terms, and adding  a quantized
topological term will always cause the system to go to another gapped phase.
So the gapped phases of a weakly coupled lattice gauge theory is classified by quantized
topological terms $H^4(BG,\Z)$ in $3$ space-time dimensions.

For finite gauge groups, the weakly coupled lattice gauge theories are also in gapped
phases which have non-trivial topological orders.  In this case, can
generalized Chern-Simons terms (\ie $H^{d+1}(BG,\Z)$) also classify the
topological phases of weakly coupled lattice gauge theories?

For finite gauge groups, we may choose $S_\text{dyn}[\{G_{ij}\}] =\infty$ for
gauge configurations with non-zero field strength, and
$S_\text{dyn}[\{G_{ij}\}] = 0$ for gauge configurations with zero field
strength.  In this case, the path integral \eqn{Zgauge} becomes topological
(\ie invariant under any coarse graining of the lattice.) Such lattice
topological gauge theories are classified by $H^{d+1}(BG,\Z)$ in $d$ space-time
dimensions.

The lattice topological gauge theories do describe gapped phases of weakly coupled
lattice gauge theories of finite gauge group.  But do those  gapped phases
belong to different phases or not?  Can those different gapped phases be
smoothly connected by deforming $S_\text{dyn}[\{G_{ij}\}]$?  Since the weakly coupled
lattice gauge theories have no global symmetries, their gapped phases all
belong to the same one phase in $2$ space-time
dimensions.\cite{VCL0501,CGW1107}  So despite that
$H^{3}(BG,\Z)$ is non-trivial
which leads to different lattice topological gauge theories, they all describe
the same phase.  However, in higher dimensions, there are non-trivial gapped
phases for  weakly coupled lattice gauge theories, and those gapped phases can be
described by the quantized topological terms $H^{d+1}(BG,\Z)$ in $d$ space-time
dimensions.  We have examples that different quantized topological terms do
give rise to topological orders, as one can see from the calculation of the
$S,T$ matrices.\cite{Wrig} However, it is not clear if
the correspondence is one-to-one for a fixed gauge group in general.

This research is supported by NSF Grant No. DMR-1005541 and NSFC 11074140.
Research at Perimeter Institute is
supported by the Government of Canada through Industry Canada and by
the Province of Ontario through the Ministry of Research. LYH is supported
by the Croucher Fellowship.

\appendix

\section{Relation between $H^*(BG,\Z)$ and $H^*(BG,\R/\Z)$ }

Since
\begin{align}
 0\to \Z \to \R \to U(1) \to 0,
\end{align}
we have
\begin{align}
... \to &
 H^0(BG,\Z)\to  H^0(BG,\R)\to  H^0(BG,\R/\Z)\to
\nonumber\\
&  H^1(BG,\Z)\to  H^1(BG,\R)\to  H^1(BG,\R/\Z)\to
\nonumber\\
&
\ \ \ \ \ \ \ \ \
\ \ \ \ \ \ \ \ \
...\ \ ...
\\
\to &  H^d(BG,\Z)\to  H^d(BG,\R)\to  H^d(BG,\R/\Z)\to ...
\nonumber
\end{align}
For a finite group $H^d(BG,\R)=0$. This allows us to obtain
\begin{align}
\label{HdHd1}
  H^d(BG,\R/\Z) \simeq H^{d+1}(BG,\Z) \simeq  \cH_B^d(G,\R/\Z) .
\end{align}
For a compact Lie group, $H^{d=\text{odd}}(BG,\R)=0$.
So we have, for $d=$ even,
\begin{align}
 0 &\to  H^{d-1}(BG,\R/\Z)\to H^d(BG,\Z)\to  H^d(BG,\R)
\nonumber\\
&\to  H^d(BG,\R/\Z)\to H^{d+1}(BG,\Z)\to  0
\end{align}
$ 0 \to  H^{d-1}(BG,\R/\Z)\to H^d(BG,\Z)$ means that each element in
$H^{d-1}(BG,\R/\Z)$ corresponds to a distinct element in $H^d(BG,\Z)$.
$H^d(BG,\R/\Z)\to H^{d+1}(BG,\Z)\to  0$ means that each  element in
$H^{d+1}(BG,\Z)$ correspond to a set
of  elements in $H^d(BG,\R/\Z)$.
So we have, for $d=$ even,
\begin{align}
\label{HRZHZ}
 H^{d-1}(BG,\R/\Z) &\subset H^d(BG,\Z)=\cH^{d-1}_B(G,\R/\Z),
\nonumber\\
H^d(BG,\R/\Z)/\Ga &= H^{d+1}(BG,\Z) =\cH^d_B(G,\R/\Z).
\end{align}
where the 0 element in $H^{d+1}(BG,\Z)$ correspond to  the subgroup $\Ga
\subset H^d(BG,\R/\Z)$.

\section{Calculate  $H^d(BG,\R/\Z)$}
\label{HBGRZ}

We can use the K\"unneth formula (see \Rf{Spa66} page 247)
\begin{align}
\label{kunn}
&\ \ \ \ H^d(X\times X',M\otimes_R M')
\nonumber\\
&\simeq \Big[\oplus_{p=0}^d H^p(X,M)\otimes_R H^{d-p}(X',M')\Big]\oplus
\nonumber\\
&\ \ \ \ \ \
\Big[\oplus_{p=0}^{d+1}
\text{Tor}_1^R(H^p(X,M),H^{d-p+1}(X',M'))\Big]  .
\end{align}
to calculate $H^*(X,M)$ from $H^*(X,Z)$.
Here $R$ is a principle ideal domain and $M,M'$ are $R$-modules
such that $\text{Tor}_1^R(M,M')=0$.
The tensor-product operation $\otimes_R$ and  the
torsion-product operation $\text{Tor}_1^R$ have the following properties:
\begin{align}
\label{tnprd}
& A \otimes_\Z B \simeq B \otimes_\Z A ,
\nonumber\\
& \Z \otimes_\Z M \simeq M \otimes_\Z \Z =M ,
\nonumber\\
& \Z_n \otimes_\Z M \simeq M \otimes_\Z \Z_n = M/nM ,
\nonumber\\
& \Z_m \otimes_\Z \Z_n  =\Z_{(m,n)} ,
\nonumber\\
&  (A\oplus B)\otimes_R M = (A \otimes_R M)\oplus (B \otimes_R M)   ,
\nonumber\\
& M \otimes_R (A\oplus B) = (M \otimes_R A)\oplus (M \otimes_R B)   ;
\end{align}
and
\begin{align}
\label{trprd}
& \text{Tor}_1^R(A,B) \simeq \text{Tor}_1^R(B,A)  ,
\nonumber\\
& \text{Tor}_1^\Z(\Z, M) = \text{Tor}_1^\Z(M, \Z) = 0,
\nonumber\\
& \text{Tor}_1^\Z(\Z_n, M) = \{m\in M| nm=0\},
\nonumber\\
& \text{Tor}_1^\Z(\Z_m, \Z_n) = \Z_{(m,n)} ,
\nonumber\\
& \text{Tor}_1^R(A\oplus B,M) = \text{Tor}_1^R(A, M)\oplus\text{Tor}_1^R(B, M),
\nonumber\\
& \text{Tor}_1^R(M,A\oplus B) = \text{Tor}_1^R(M,A)\oplus\text{Tor}_1^R(M,B)
,
\end{align}
where $(m,n)$ is the greatest common divisor of $m$ and $n$.

If we choose $R=M=\Z$, then the condition
$\text{Tor}_1^R(M,M')=\text{Tor}_1^{\Z}(\Z,M')=0$ is always satisfied. So we
have
\begin{align}
\label{kunnZ}
&\ \ \ \ H^d(X\times X',M')
\nonumber\\
&\simeq \Big[\oplus_{p=0}^d H^p(X,\Z)\otimes_{\Z} H^{d-p}(X',M')\Big]\oplus
\nonumber\\
&\ \ \ \ \ \
\Big[\oplus_{p=0}^{d+1}
\text{Tor}_1^{\Z}(H^p(X,{\Z}),H^{d-p+1}(X',M'))\Big]  .
\end{align}
Now we can further choose $X'$ to be the space of one point, and use
\begin{align}
H^{d}(X',M'))=
\begin{cases}
M', & \text{ if } d=0,\\
0, & \text{ if } d>0,
\end{cases}
\end{align}
to reduce \eqn{kunnZ} to
\begin{align}
\label{ucf}
&\ \ \ \ H^d(X,M)
\\
&\simeq  H^d(X,\Z)\otimes_{\Z} M \oplus
\text{Tor}_1^{\Z}(H^{d+1}(X,{\Z}),M)  ,
\nonumber
\end{align}
where $M'$ is renamed as $M$.  The above is a form of the universal coefficient
theorem which can be used to calculate $H^*(BG,M)$ from $H^*(BG,\Z)$ and the
module $M$.

Now, let us choose $M=\R/\Z$. Note that $H^d (BG,\Z)$ has a form $H^d
(BG,\Z)=\Z\oplus ... \oplus \Z \oplus Z_{n_1} \oplus Z_{n_2}\oplus ...  $.  A
$\Z$ in $H^d (BG,\Z)$ will produce a $\R/\Z$ in $H^d (BG,\R/\Z)$ since $\Z
\otimes_{\Z} \R/\Z=\R/\Z$.  A $\Z_n$ in $H^{d+1} (BG,\Z)$ will produce a $\Z_n$
in $H^d (BG,\R/\Z)$ since $\text{Tor}_1^{\Z}(\Z_n, \R/\Z)=\Z_n$.  So we see
that $H^d (BG,\R/\Z)$ has a form $H^d (BG,\R/\Z)=\R/\Z\oplus ... \oplus \R/\Z
\oplus Z_{n_1} \oplus Z_{n_2}\oplus ...  $ and
\begin{align}
\label{DisH}
\text{Dis}[ H^d (X,\R/\Z) ] \simeq \text{Tor}[H^{d+1}(X,\Z)] .
\end{align}
where $\text{Dis}[ H^d (X,\R/\Z) ]$ is the discrete part of $H^d (X,\R/\Z)$.

If we choose $M=\R$, we find that
\begin{align}
\label{ucfR}
 H^d(X,\R)
\simeq  H^d(X,\Z)\otimes_{\Z} \R.
\end{align}
So $ H^d(X,\R)$ has the form $\R\oplus ... \oplus \R$ and each $\Z$ in
$H^d(X,\Z)$ gives rise to a $\R$ in $ H^d(X,\R)$.
Since $ H^d(BG,\R)=0$ for $d =$ odd,
we have
\begin{align}
\label{HTorH}
  H^d(BG,\Z)=\text{Tor}[ H^d(BG,\Z)],\ \ \text{ for } d= \text{ odd}.
\end{align}

\bibliographystyle{apsrev}

\end{document}